\documentclass[twocolumn]{aastex62}

\usepackage{color}

\graphicspath{{./}{figures/}}

\received{August 13, 2018}
\revised{September 7, 2018}
\accepted{September 20, 2018}

\submitjournal{ApJ}

\shorttitle{Filaments around Abell 133}
\shortauthors{Connor et al.}

\begin{document}

\title{Wide-Field Optical Spectroscopy of Abell 133: A Search for Filaments Reported in X-ray Observations}

\correspondingauthor{Thomas Connor}
\email{tconnor@carnegiescience.edu}

\author[0000-0002-7898-7664]{Thomas Connor}
\affil{The Observatories of the Carnegie Institution for Science, 813 Santa Barbara St., Pasadena, CA 91101, USA}

\author[0000-0003-4727-4327]{Daniel D. Kelson}
\affiliation{The Observatories of the Carnegie Institution for Science, 813 Santa Barbara St., Pasadena, CA 91101, USA}

\author[0000-0003-2083-5569]{John Mulchaey}
\affiliation{The Observatories of the Carnegie Institution for Science, 813 Santa Barbara St., Pasadena, CA 91101, USA}

\author[0000-0001-8121-0234]{Alexey Vikhlinin}
\affil{Harvard-Smithsonian Center for Astrophysics, 60 Garden Street, Cambridge, MA 02138, USA}

\author{Shannon G. Patel}
\affiliation{The Observatories of the Carnegie Institution for Science, 813 Santa Barbara St., Pasadena, CA 91101, USA}

\author[0000-0003-4849-9536]{Michael L. Balogh}
\affiliation{Department of Physics and Astronomy, University of Waterloo, Waterloo, Ontario, ON N2L 3G1, Canada}

\author{Gandhali Joshi}
\affiliation{Max-Planck-Institut f\"{u}r Astronomie, K\"{o}nigstuhl 17, D-69117 Heidelberg, Germany}
\affiliation{Department of Physics and Astronomy, University of Waterloo, Waterloo, Ontario, ON N2L 3G1, Canada}

\author[0000-0002-0765-0511]{Ralph Kraft}
\affiliation{Harvard-Smithsonian Center for Astrophysics, 60 Garden Street, Cambridge, MA 02138, USA}

\author[0000-0002-6766-5942]{Daisuke Nagai}
\affiliation{Department of Physics, Yale University, New Haven, CT 06520, USA}
\affiliation{Yale Center for Astronomy and Astrophysics, Yale University, New Haven, CT 06520, USA}

\author{Svetlana Starikova}
\affiliation{Harvard-Smithsonian Center for Astrophysics, 60 Garden Street, Cambridge, MA 02138, USA}

\begin{abstract}
Filaments of the cosmic web have long been associated with the threadlike structures seen in galaxy redshift surveys. However, despite their baryon content being dominated by hot gas, these filaments have been an elusive target for X-ray observations. Recently, detections of filaments in very deep (2.4 Msec) observations with \textit{Chandra} were reported around Abell 133 ($z=0.0559$). To verify these claims, we conducted a multi-object spectrographic campaign on the Baade 6.5m telescope around Abell 133; this resulted in a catalog of ${\sim}3000$ new redshift measurements, of which 254 are of galaxies near the cluster. We investigate the kinematic state of Abell 133 and identify the physical locations of filamentary structure in the galaxy distribution. Contrary to previous studies, we see no evidence that Abell 133 is dynamically disturbed; we reject the hypothesis that there is a kinematically distinct subgroup ($p=0.28$) and find no velocity offset between the central galaxy and the cluster ($\textrm{Z}_\textrm{score}=0.041^{+0.111}_{-0.106}$).  The spatial distribution of galaxies traces the X-ray filaments, as confirmed by angular cross correlation with a significance of ${\sim}5\sigma$. A similar agreement is found in the angular density distribution, where two X-ray structures have corresponding galaxy enhancements. We also identify filaments in the large-scale structure of galaxies; these filaments approach the cluster from the direction the X-ray structures are seen. While more members between $\textrm{R}_{200}$ and $2\times\textrm{R}_{200}$ are required to clarify which large scale filaments connect to the X-ray gas, we argue that this is compelling evidence that the X-ray emission is indeed associated with cosmic filaments.
\end{abstract}


\section{Introduction} \label{sec:intro}

One of the defining characteristics of the large-scale structure of the Universe is its filamentary nature, where galaxy clusters are at nodes linked by threads of dark matter and baryons. Dating back to the work of \citet{1983MNRAS.204..891K}, this pattern, commonly referred to as the cosmic web \citep{1996Natur.380..603B}, has been an ever-present result of cosmological simulations \citep[e.g.,][]{1987ApJ...313..505W,2005Natur.435..629S,2009MNRAS.398.1150B,2014MNRAS.444.1518V}. It can also be seen as a prediction of fundamental theories of structure formation \citep[e.g.,][]{1970A&A.....5...84Z,1980MNRAS.192..321D,1982Natur.300..407Z,1993MNRAS.260..765C,1995ApJ...439..520E}, whereby matter disperses out of voids into sheets and filaments, which feed onto clusters \citep{1999MNRAS.308..593C,2005MNRAS.359..272C,2012MNRAS.425.2049H,2014MNRAS.441.2923C}. Models predict that 2-5 filaments connect to a typical cluster \citep{2010MNRAS.408.2163A} and these filaments are expected to be ${\sim} 1\ {\rm Mpc}$ thick. Galaxies in filaments provide ${\sim}40\%$ of the total galaxy luminosity \citep{2014MNRAS.438.3465T} and gas in filaments contain a similar fraction of the total baryonic mass of the Universe \citep{2001ApJ...552..473D}.

Filaments have long been seen in the statistical distribution of galaxies, even before the advent of large-scale redshift surveys \citep{1978MNRAS.185..357J, 1978ApJ...222..784G}. Larger spectroscopic surveys, such as the CfA redshift survey \citep{1986ApJ...302L...1D, 1989Sci...246..897G}, the Las Campanas Redshift Survey \citep{1996ApJ...470..172S}, and the 2dF survey \citep{2001MNRAS.328.1039C}, revealed the nature of the cosmic web, and modern programs continue to map out the distribution of galaxies across the Universe \citep[e.g.,][]{2009MNRAS.399..683J, 2013AJ....145...10D, 2014A&A...566A.108G}.
Closer in to clusters, filaments are detectable in weak lensing maps \citep{2010MNRAS.401.2257M}, and a number of studies have reported detecting dark matter filaments this way \citep{2011A&A...532A..57S,2012Natur.487..202D,2012MNRAS.426.3369J,2015Natur.528..105E,2015arXiv150306373H}. Filaments can also be identified through the properties of their constituent galaxies \citep{2008ApJ...672L...9F,2015A&A...576L...5T, 2016MNRAS.455.2267R}. 

While galaxies are useful for tracing the positions of filaments, the baryonic gas that makes up the filaments is even more intriguing. Around one-half of the baryons in the low-redshift Universe are expected to reside in these filaments in the form of warm-hot ($10^5 - 10^7$ K) intergalactic medium \citep[WHIM;][]{1999ApJ...514....1C, 2001ApJ...552..473D}. This gas is believed to have been shock heated as it gravitationally assembles \citep{2002A&A...388..741V}. The gaseous component of filaments has been seen in Lyman-$\alpha$ emission \citep{2014Natur.506...63C} and the Sunyaev-Zel'dovich effect \citep{2017arXiv170910378D,2018A&A...609A..49B}, and the gas properties have been measured with radio observations \citep{2010ApJ...724L.143E}. But, despite the expectation for X-ray emission from the WHIM, it is difficult to observe due to the low signal relative to the soft, diffuse X-ray background \citep{2009ApJ...695.1127G}.

The first dedicated search for filaments in X-rays was by \citet{1995A&A...302L...9B}, who were unable to detect any filaments in the {\it ROSAT} All-Sky Survey. Since then, the search for X-ray detection of filaments has been full of non-detections and potential structures lacking statistical significance.
\citet{1999A&A...341...23K} saw enhanced X-ray emission between clusters, although not outside of their projected virial regions. \citet{2000ApJ...528L..73S} saw evidence for a low-density structure, but their detection occurred in a region of high Galactic background. 
\citet{2006ApJ...652..189K} showed that a previously reported detection of filamentary absorption by \citet{2005ApJ...629..700N} was not statistically significant. Similarly, a report of absorption by \citet{2003ApJ...582...82M} was only at a 2$\sigma$ level. 
\citet{2002ApJ...572L.127F} reported \ion{O}{8} absorption from cosmic filaments, but further analysis by \citet{2003NewAR..47..561C} could not confirm those claims. 
\citet{2003A&A...397..445K} and \citet{2003A&A...410..777F} claimed detections of soft X-ray emission from filaments, but they lacked the energy resolution to rule out contributions from the Galactic foreground \citep{2003PASJ...55..879Y}.
A report of an X-ray filament along Abell 85 \citep{2003A&A...403L..29D, 2005A&A...432..809D} was shown by optical spectroscopy to be more accurately considered an infalling chain of groups \citep{2008A&A...489...11B}. 
\citet{2012PASJ...64...18M} were unable to detect filamentary absorption in a region that shows secondary indicators of being a filament \citep{2015ApJ...806..211U}. 

Part of the reason for the trouble in detecting X-ray emission from filaments outside of the virial radius is the ubiquity of the X-ray background; \citet{2001ApJ...548L.123V} showed that as much as ${\sim}1/3$ of the X-ray sky is covered by the virialized regions of groups and clusters. The other main problem is that X-ray filaments are rather diffuse and, since their emission is proportional to their density squared, rather faint \citep[e.g.,][]{2015MNRAS.453.1164G}.
\begin{figure*}
\plotone{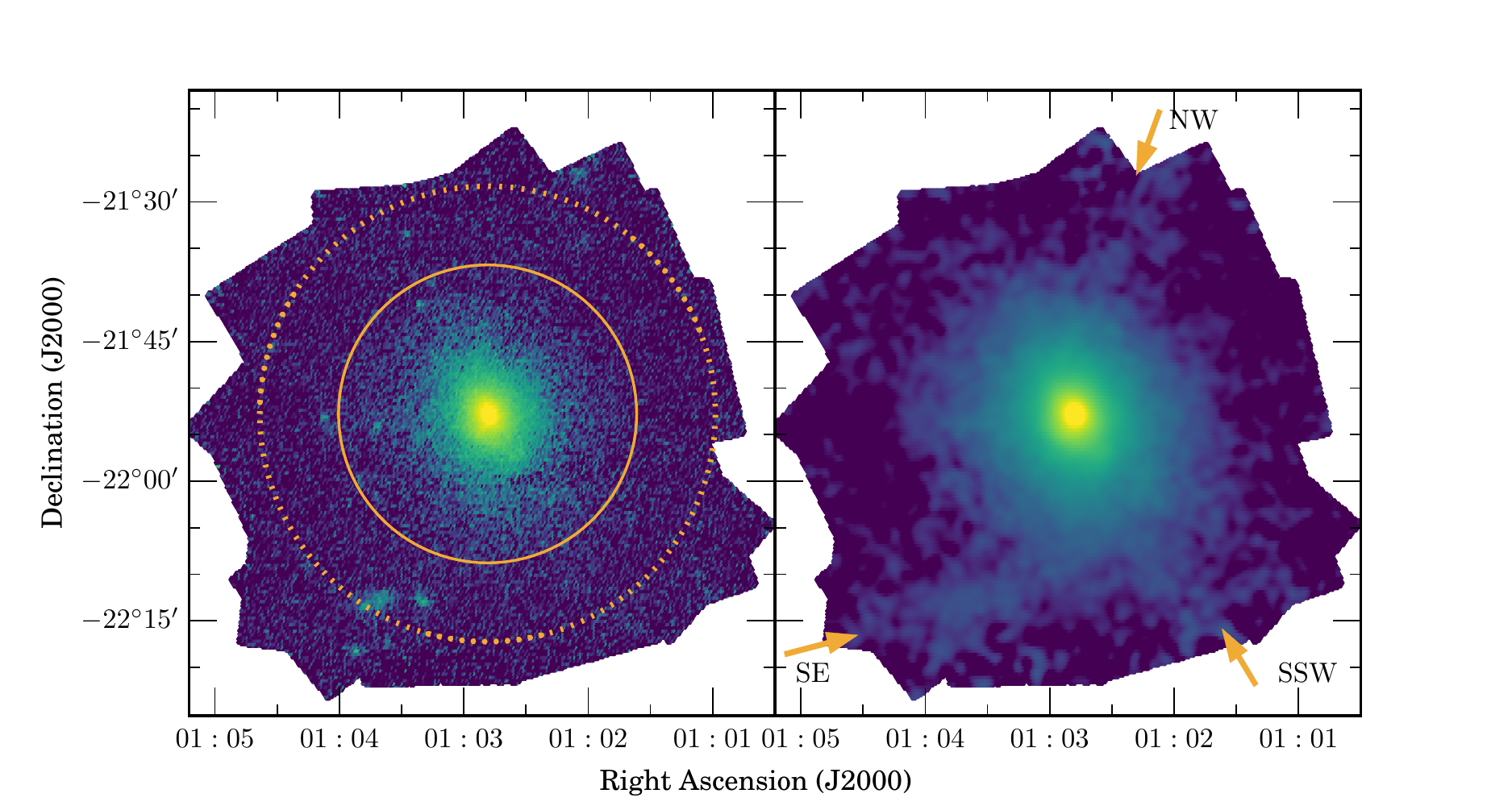}
\caption{Combined {\it Chandra} X-ray observations of Abell 133. {\bf Left:} Combined image after subtraction of point sources. Intensity is shown with logarithmic scaling. The solid orange circle traces $R_{500}$ (1044 kpc), while the dashed orange circle traces $R_{200}$ (1596 kpc). {\bf Right:} X-ray map after removal of wavelet-detected structures. This image has been smoothed by a Gaussian kernel. Filaments identified by \citet{2013HEAD...1340101V} are marked with orange arrows and labeled. The overall cluster intensity profile is still included, but the effects of individual clumps have been mitigated. In both images, the white areas correspond to areas without any {\it Chandra} coverage.}. \label{fig:two_xray_images}
\end{figure*}

One way to maximize the chance of detecting X-ray emission from filaments is to look for filaments near merging and neighboring clusters, where the increase in density and temperature decrease the difficulty of observation. \citet{2008A&A...482L..29W} detected an excess of X-ray emission connecting Abell 222 and Abell 223 coincident with an observed dark matter filament \citep{2012Natur.487..202D}. This filament is oriented into the plane of the sky, maximizing the surface density and thereby increasing the observed emission but never extending beyond the projected virial radii \citep{2005A&A...440..453D}. 
\citet{2001ApJ...563..673T} saw evidence of an X-ray filament between Abell 3391 and Abell 3395 \citep[see also][]{2017PASJ...69...93S,2018ApJ...858...44A}; again, the 
two clusters may be merging and the filament is only seen inside the virial radii \citep{2017A&A...606A...1A}. 
\citet{2016ApJ...818..131B} identified a filament in X-ray images of Abell 1750, connecting three individual sub-clusters. This system is most likely a triple-merger system, and the filament is only seen along the triple-merger axis.
 While these detections represent a marked increase from the pioneering efforts of \citet{1995A&A...302L...9B}, they provide a biased look at cosmic filaments. If detection is made easier by the density and temperature being increased through a merger, then filaments between merging clusters will always sample the hottest and densest filaments. The outstanding task is X-ray observations of more cosmically normal filaments around a cluster.

Recently, \citet{2015Natur.528..105E} reported detecting filaments in a ${\sim}100$ ksec observation of Abell 2744 with {\it XMM-Newton}, which also appear in weak lensing maps. Follow-up observations of this cluster using {\it Suzaku} by \citet{2017PASJ...69...39H} found that the X-ray spectrum in the filamentary regions was better fit when including a WHIM emission model, but despite having ${\sim}300$ total ksec of observations, they did not detect any structure with any statistical significance. While these filaments do not link merging clusters, Abell 2744 appears to currently be in the late stages of a major merger \citep{2004MNRAS.349..385K, 2011ApJ...728...27O, 2011MNRAS.417..333M, 2018MNRAS.473..663M}. Strong lensing analysis of Abell 2744 has revealed an extreme amount of substructure in the cluster \citep{2016MNRAS.463.3876J}, which \citet{2017MNRAS.467.2913S} showed is at the limit allowable in standard $\Lambda$CDM cosmology. 

\begin{deluxetable*}{rrrrrrr}
\tablecaption{{\it Chandra} Observations}
\label{tab:CXOobs}
\tablewidth{0pt}
\tablehead{
\colhead{ID} & \colhead{Obs ID} & \colhead{Exposure Time} & \colhead{$\alpha_{2000}$} & \colhead{$\delta_{2000}$} & \multicolumn2c{Start Time (UTC)}\\
\colhead{} & \colhead{} & \colhead{(ks)} & \colhead{} & \colhead{} & \colhead{(YYYY-mm-dd)} & \colhead{(hh:mm:ss)}}
\startdata
1  & 3183  &  44.52 & $01^h 02^m 40\fs80$ & $-21\arcdeg 52\arcmin 40\farcs80$ &	2002-06-24 & 06:33:56\\
2  & 3710  &  44.59 & $01^h 02^m 40\fs80$ & $-21\arcdeg 52\arcmin 40\farcs80$ &	2002-06-26 & 15:50:06\\
3  & 9410  &  19.91 & $01^h 05^m 37\fs13$ & $-24\arcdeg 40\arcmin 49\farcs70$ &	2008-08-11 & 15:47:18\\
4  & 9897  &  69.22 & $01^h 02^m 41\fs80$ & $-21\arcdeg 52\arcmin 50\farcs00$ &	2008-08-29 & 16:24:07\\
5  & 12177 &  50.11 & $01^h 01^m 48\fs65$ & $-22\arcdeg 03\arcmin 41\farcs44$ &	2010-08-31 & 19:36:10\\
6  & 12178 &  46.83 & $01^h 03^m 59\fs79$ & $-22\arcdeg 00\arcmin 07\farcs78$ &	2010-09-07 & 04:51:24\\
7  & 12179 &  51.10 & $01^h 02^m 56\fs40$ & $-22\arcdeg 08\arcmin 25\farcs87$ &	2010-09-03 & 13:20:16\\
8  & 13391 &  46.43 & $01^h 03^m 01\fs71$ & $-22\arcdeg 13\arcmin 59\farcs05$ &	2011-08-16 & 06:07:24\\
9  & 13392 &  49.89 & $01^h 02^m 33\fs41$ & $-21\arcdeg 33\arcmin 46\farcs36$ &	2011-09-16 & 01:54:49\\
10 & 13442 & 176.69 & $01^h 01^m 24\fs06$ & $-21\arcdeg 49\arcmin 03\farcs25$ &	2011-08-23 & 07:02:36\\
11 & 13443 &  69.68 & $01^h 01^m 26\fs05$ & $-21\arcdeg 48\arcmin 43\farcs63$ &	2011-08-26 & 12:28:41\\
12 & 13444 &  38.26 & $01^h 03^m 05\fs78$ & $-21\arcdeg 36\arcmin 48\farcs28$ &	2011-09-03 & 16:58:43\\
13 & 13445 &  65.18 & $01^h 02^m 03\fs25$ & $-22\arcdeg 11\arcmin 34\farcs15$ &	2011-09-02 & 21:57:03\\
14 & 14333 & 134.76 & $01^h 03^m 05\fs78$ & $-21\arcdeg 36\arcmin 48\farcs28$ &	2011-08-31 & 05:33:29\\
15 & 13446 &  58.42 & $01^h 01^m 55\fs23$ & $-21\arcdeg 35\arcmin 13\farcs50$ &	2011-09-09 & 15:26:18\\
16 & 13447 &  69.13 & $01^h 03^m 55\fs81$ & $-21\arcdeg 44\arcmin 39\farcs69$ &	2011-09-08 & 19:39:59\\
17 & 14338 & 117.50 & $01^h 01^m 55\fs23$ & $-21\arcdeg 35\arcmin 13\farcs50$ &	2011-09-10 & 20:32:51\\
18 & 13448 & 146.12 & $01^h 03^m 58\fs33$ & $-21\arcdeg 42\arcmin 37\farcs60$ &	2011-09-13 & 09:56:39\\
19 & 13449 &  68.15 & $01^h 01^m 41\fs65$ & $-21\arcdeg 39\arcmin 44\farcs81$ &	2011-09-06 & 03:23:37\\
20 & 14343 &  35.30 & $01^h 03^m 58\fs33$ & $-21\arcdeg 42\arcmin 37\farcs60$ &	2011-09-12 & 06:08:44\\
21 & 13451 &  70.12 & $01^h 03^m 47\fs33$ & $-22\arcdeg 11\arcmin 44\farcs50$ &	2011-09-16 & 20:11:26\\
22 & 13452 & 142.07 & $01^h 01^m 17\fs20$ & $-22\arcdeg 05\arcmin 22\farcs14$ &	2011-09-24 & 12:29:41\\
23 & 14345 &  33.74 & $01^h 01^m 17\fs20$ & $-22\arcdeg 05\arcmin 22\farcs14$ &	2011-09-23 & 03:35:49\\
24 & 13454 &  91.79 & $01^h 04^m 07\fs28$ & $-21\arcdeg 53\arcmin 49\farcs43$ &	2011-09-19 & 19:46:28\\
25 & 14346 &  85.90 & $01^h 04^m 07\fs28$ & $-21\arcdeg 53\arcmin 49\farcs43$ &	2011-09-21 & 19:44:45\\
26 & 13456 & 135.63 & $01^h 02^m 32\fs52$ & $-22\arcdeg 08\arcmin 52\farcs11$ &	2011-10-15 & 15:20:26\\
27 & 14354 &  38.64 & $01^h 02^m 32\fs52$ & $-22\arcdeg 08\arcmin 52\farcs11$ &	2011-10-10 & 17:52:58\\
28 & 13450 & 108.18 & $01^h 03^m 54\fs12$ & $-22\arcdeg 09\arcmin 38\farcs37$ &	2011-10-05 & 02:18:20\\
29 & 14347 &  68.69 & $01^h 03^m 54\fs12$ & $-22\arcdeg 09\arcmin 38\farcs37$ &	2011-10-09 & 22:02:04\\
30 & 13453 &  68.95 & $01^h 03^m 21\fs95$ & $-21\arcdeg 36\arcmin 15\farcs65$ &	2011-10-13 & 22:40:53\\
31 & 13455 &  69.63 & $01^h 03^m 55\fs97$ & $-21\arcdeg 54\arcmin 56\farcs68$ &	2011-10-19 & 01:50:46\\
32 & 13457 &  69.13 & $01^h 01^m 21\fs73$ & $-22\arcdeg 05\arcmin 10\farcs47$ &	2011-10-21 & 17:22:13\\
\enddata
\end{deluxetable*}

Previously, \citet{2013HEAD...1340101V} reported the detection of three filaments in an extremely deep observation of Abell 133, shown in Figure \ref{fig:two_xray_images}. While the X-ray analysis has not yet been published (Vikhlinin et al., in prep), the announced detection of filaments around this cluster has been considered in the development of future X-ray missions. Abell 133 \citep[also known as RXC J0102.7$-$2152]{1958ApJS....3..211A} is a cool-core cluster \citep[$K_0 = 17.26\ {\rm keV}\ {\rm cm}^2$,][]{2009ApJS..182...12C} with a cooling flow first identified by \citet{1997MNRAS.292..419W}. Early {\it Chandra} \citep{2002ApJ...575..764F} and {\it XMM-Newton} \citep{2004ApJ...616..157F} observations showed that the cluster core has a complex morphology. \citet{2001AJ....122.1172S} reported the detection of a radio relic approximately $30\arcsec$ N of the cluster center. Previous analysis of this cluster by \citet{2006ApJ...640..691V} found values\footnote{$M_{\Delta}$ and $R_{\Delta}$ (where, in this work, $\Delta = 200$, $500$, or $2500$) is the enclosed mass and radius, respectively, of a sphere with a density $\rho = \Delta \times \rho_c(z)$, where $\rho_c(z)$ is the critical density of the Universe at redshift z.} of $M_{2500} = 1.13 \pm 0.07 \times 10^{14}\ {\rm M}_\odot$ and $M_{500} = 3.17 \pm 0.38 \times 10^{14}\ {\rm M}_\odot$. Based on that mass measurement, the expectation is for Abell 133 to have $3 \pm 1$ connecting filaments \citep{2010MNRAS.408.2163A}, which is in line with what was announced by \citet{2013HEAD...1340101V}. Not only is Abell 133 possibly the closest to an ideal system for studying the filaments of the cosmic web yet seen,but the depth of the {\it Chandra} observations tentatively give us the cleanest view of any X-ray filaments out to and beyond the virial radius.

One concern for the interpretation of possible filaments is the presence of substructure in Abell 133. \citet{1999A&A...351..883K} noticed an enhancement of galaxy signal to the SW of the cluster center in a mapping of individual galaxies. A spectroscopic analysis of the cluster by \citet{1997astro.ph..9036W} was inconclusive, with a 2 dimensional analysis finding evidence not supported in a 3 dimensional analysis, although they did find statistical support for a high-velocity cD galaxy. \citet{2006A&A...450....9F} identified two substructures in a 2 dimensional wavelet-based analysis; the one previously identified by \citet{1999A&A...351..883K} as well as a smaller clump to the NE of the cluster center. Combining galaxy positions, X-ray observations of the cluster center, and radio data, \citet{2010ApJ...722..825R} proposed that Abell 133 has recently experienced a merger, whereby an infalling subcluster passed from NE to SW. An understanding of the cluster's kinematic state is needed to discern between filaments and merging subclusters.

In this work, we describe the results of a campaign of optical spectroscopy performed on the Magellan-Baade telescope to study Abell 133. With ${\sim}3000$ spectroscopic redshifts, of which ${\sim}250$ are for galaxies at the velocity of the cluster, we diagnose the kinematic state of Abell 133, finding no statistically significant evidence of substructure. In combination with archival DECam photometry of the surrounding area, we show that the distribution of galaxies matches the distribution of X-ray emission, at least for two of the filaments. At large scales, the filamentary network of galaxies intercepts the cluster where expected by the {\it Chandra} predictions. We first present the X-ray observations used in this work (although their analysis is left to a later paper) in Section \ref{sec:xray}. In Section \ref{sec:optical} we describe the optical observations undertaken and the analysis procedures used to determine redshifts. In Sections \ref{sec:galaxies} and \ref{sec:CosmicWeb} we discuss how the X-ray observations compare to optical observations of galaxies around the X-ray footprint and in the cosmic web, respectively. Finally, in Section \ref{sec:Discussion}, we discuss our results in the context of the cosmic web and the outlook of future studies.
Throughout this work, we assume a flat $\Lambda$CDM cosmology with $\Omega_M = 0.3$, and ${\rm H}_0 = 70\ {\rm km}\ {\rm s}^{-1}\ {\rm Mpc}^{-1}$. We adopt a nominal redshift for Abell 133 of $z_c = 0.0559$ and a central position of $\alpha_{2000} = 01^h 02^m 41\fs91$, $\delta_{2000} = -21\arcdeg 52\arcmin 51\farcs5$. At this redshift, $1\arcsec = 1.085\ {\rm kpc}$. We computed relative velocities using the formula 
\begin{equation} \label{eqn:vel_calc}
v = \frac{z - z_c}{1 + z_c} \times c,
\end{equation}
where c is the speed of light.

\section{X-Ray Observations} \label{sec:xray}

\begin{deluxetable*}{rrllrrlrrr}
\tablecaption{IMACS Observing Log}
\label{tab:IMACS}
\tablewidth{0pt}
\tablehead{
\colhead{ID} & \colhead{Name} & \colhead{$\alpha_{2000}$} & \colhead{$\delta_{2000}$} & \colhead{Date\tablenotemark{a}} & \colhead{Exposure Time} & \colhead{Seeing\tablenotemark{b}} & \colhead{${\rm N}_{\rm Obj}$}& \colhead{${\rm N}_{\rm C}$\tablenotemark{c}} & \colhead{${\rm N}_{\rm U}$}\\
\colhead{} & \colhead{} & \colhead{} & \colhead{} & \colhead{(YYYY-mm-dd)} & \colhead{(s)} & \colhead{(arcsec)} & \colhead{} & \colhead{} & \colhead{}}
\startdata
0 & MA1 & $01^h 03^m 34\fs1$ & $-22\arcdeg 06\arcmin 00\farcs0$ & 2011-09-28 & $2\times1800$ & $0\farcs8$ & 89 & 21 & 1 \\
1 & MB1 & $01^h 01^m 43\fs7$ & $-22\arcdeg 09\arcmin 08\farcs8$ & 2011-09-29 & $2\times900$ & $0\farcs45$ & 91 & 25 & 10 \\
2 & MA2 & $01^h 03^m 34\fs1$ & $-22\arcdeg 06\arcmin 00\farcs0$ & 2011-09-30 & $2\times1800$ & $0\farcs8$ & 83 & 15 & 2 \\
3 & MB2 & $01^h 01^m 47\fs5$ & $-22\arcdeg 09\arcmin 08\farcs8$ & 2011-10-01 & $2\times1800$ & \nodata & 82 & 25 & 15 \\
4 & MA3 & $01^h 03^m 34\fs1$ & $-22\arcdeg 02\arcmin 50\farcs0$ & 2011-10-01 & $2\times900$ & $0\farcs8$ & 67 & 2 & 18 \\
5 & M1 & $01^h 03^m 36\fs8$ & $-22\arcdeg 06\arcmin 00\farcs0$ & 2013-09-07 & $3\times1800$ & $1\farcs1$ & 134 & 16 & 22 \\
6 & M2 & $01^h 03^m 36\fs8$ & $-22\arcdeg 12\arcmin 45\farcs0$ & 2013-09-07 & $2\times1800$ & $0\farcs7$ & 137 & 9 & 13 \\
7 & M3 & $01^h 01^m 27\fs1$ & $-22\arcdeg 06\arcmin 00\farcs0$ & 2013-09-07 & $2\times1800$ & $0\farcs6$ & 123 & 21 & 14 \\
8 & M4 & $01^h 01^m 26\fs5$ & $-22\arcdeg 14\arcmin 00\farcs0$ & 2013-09-07 & $2\times1800$ & $0\farcs55$ & 128 & 15 & 14 \\
9 & M5 & $01^h 01^m 26\fs5$ & $-21\arcdeg 54\arcmin 30\farcs0$ & 2013-09-07 & $2\times1800$ & $1\farcs1$ & 126 & 8 & 24 \\
10 & M6 & $01^h 01^m 30\fs0$ & $-21\arcdeg 44\arcmin 45\farcs0$ & 2013-09-09 & $2\times1800$ & $0\farcs75$ & 137 & 8 & 9 \\
11 & M7 & $01^h 01^m 30\fs0$ & $-21\arcdeg 34\arcmin 30\farcs0$ & 2013-09-09 & $3\times1800$ & $0\farcs8$ & 136 & 6 & 19 \\
12 & M8 & $01^h 01^m 30\fs0$ & $-21\arcdeg 24\arcmin 00\farcs0$ & 2013-09-09 & $2\times1800$ & $0\farcs9$ & 141 & 8 & 11 \\
13 & M9 & $01^h 03^m 36\fs8$ & $-21\arcdeg 24\arcmin 50\farcs0$ & 2013-09-09 & $2\times1800$ & $0\farcs7$ & 136 & 10 & 13 \\
14 & M10 & $01^h 03^m 31\fs0$ & $-21\arcdeg 34\arcmin 00\farcs0$ & 2013-09-09 & $2\times1800$ & $0\farcs7$ & 133 & 12 & 9 \\
15 & M11B & $01^h 03^m 30\fs1$ & $-21\arcdeg 45\arcmin 12\farcs4$ & 2014-09-25 & $2\times1800$ & $0\farcs6$ & 136 & 12 & 23 \\
16 & M12 & $01^h 03^m 31\fs0$ & $-21\arcdeg 54\arcmin 30\farcs0$ & 2014-07-07 & $2\times1800$ & $0\farcs65$ & 123 & 15 & 24 \\
17 & M13 & $01^h 04^m 08\fs0$ & $-22\arcdeg 09\arcmin 00\farcs0$ & 2014-07-07 & $2\times1800$ & $0\farcs6$ & 133 & 1 & 17 \\
18 & M14 & $01^h 03^m 00\fs0$ & $-22\arcdeg 09\arcmin 00\farcs0$ & 2014-07-08 & $2\times1200$ & $0\farcs5$ & 151 & 7 & 27 \\
19 & M15 & $01^h 01^m 52\fs0$ & $-22\arcdeg 09\arcmin 00\farcs0$ & 2014-07-08 & $2\times1200$ & $0\farcs5$ & 138 & 13 & 15 \\
20 & M16 & $01^h 01^m 03\fs0$ & $-21\arcdeg 47\arcmin 30\farcs0$ & 2014-09-25 & $2\times1800$ & \nodata & 132 & 8 & 13 \\
21 & M17 & $01^h 02^m 19\fs0$ & $-21\arcdeg 30\arcmin 30\farcs0$ & 2014-09-25 & $2\times1800$ & $0\farcs77$ & 144 & 2 & 21 \\
22 & M18 & $01^h 01^m 38\fs5$ & $-21\arcdeg 37\arcmin 15\farcs0$ & 2014-09-25 & $2\times1800$ & $0\farcs65$ & 146 & 1 & 10 \\
23 & M19 & $01^h 03^m 45\fs0$ & $-21\arcdeg 46\arcmin 00\farcs0$ & 2014-09-25 & $2\times1800$ & $0\farcs64$ & 141 & 3 & 10 \\
24 & M20 & $01^h 03^m 18\fs0$ & $-21\arcdeg 27\arcmin 00\farcs0$ & 2014-09-26 & $2\times1800$ & $0\farcs73$ & 148 & 1 & 30 \\
25 & M21 & $01^h 02^m 48\fs0$ & $-21\arcdeg 52\arcmin 00\farcs0$ & 2014-09-26 & $2\times1800$ & $1\farcs0$ & 133 & 12 & 37 \\
26 & M22 & $01^h 01^m 31\fs5$ & $-22\arcdeg 09\arcmin 20\farcs0$ & 2014-09-26 & $2\times1800$ & $0\farcs8$ & 147 & 3 & 43 \\
27 & M24 & $01^h 04^m 17\fs0$ & $-21\arcdeg 29\arcmin 00\farcs0$ & 2014-09-26 & $2\times1800$ & $0\farcs85$ & 125 & 2 & 14 \\
28 & M25 & $01^h 04^m 26\fs0$ & $-21\arcdeg 54\arcmin 00\farcs0$ & 2014-09-26 & $2\times1800$ & \nodata & 140 & 1 & 19 \\
\enddata
\tablenotetext{a}{Date at start of the night}
\tablenotetext{b}{Worst seeing reported by the observer for those observations}
\tablenotetext{c}{Number of cluster members, defined as galaxies with velocities within $2500\ {\rm km}\ {\rm s}^{-1}$ of the cluster redshift.}
\end{deluxetable*}

Abell 133 was observed by the {\it Chandra X-Ray Observatory} (CXO) as part of X-ray Visionary Project 13800509 (PI: Vikhlinin) and for GTO Proposals 03800034, 10800035, and 12800064 (PIs: Murray). The combined images are shown in Figure \ref{fig:two_xray_images}. All images were taken with ACIS-I in VFAINT data mode and the Timed Exposure (TE) exposure mode. Details of the observations used in this work are given in Table \ref{tab:CXOobs}. For the observations from program 13800509, the entirety of the Abell 133 field was observed to a depth of ${\sim}250$ ksec. The details of the reduction and analysis of these observations will be presented in Vikhlinin et al. (in prep). The relevant results from that work, as well as from \citet{2013HEAD...1340101V}, are summarized below, but the important result in the context of this study is the detection of the three faint X-ray structures identified in Figure \ref{fig:two_xray_images}.

Figure \ref{fig:two_xray_images} shows the combined X-ray observations first presented by \citet{2013HEAD...1340101V}. On the left, we show the image after point source removal, with a few handfuls of clumps visible. On the right, we show the image after subtracting off these clumps and performing a Gaussian smoothing of the image. \citet{2013HEAD...1340101V} identified three filaments in this image, which are marked by orange arrows in Figure \ref{fig:two_xray_images}: one in the SE, one in the SSW, and one in the NW. These filaments have X-ray temperatures ${\rm T}_{\rm X} \sim 2\ {\rm keV}$, gas contents of ${\rm M}_{\rm Gas} / {\rm M}_{\rm Tot} \sim 0.2$, and widths of $r_c \sim 0.5\ {\rm Mpc}$. We note that the X-ray coverage extends up to $R_{200}$ in all directions and beyond that radius in some directions, notably around the structure identified as the SE filament.


These observations have also been previously studied by \citet{2014MNRAS.437.1909M}. While the goal of their work was to study the clumping fraction of Abell 133 with respect to radius, they also analyzed the high-frequency components of the X-ray data to identify residual structure. \citet{2014MNRAS.437.1909M} identified an excess of structure in two funnel-like projections, covering position angles of $0^\circ$ to $45^\circ$ and $180^\circ$ to $225^\circ$ measured E of N. These funnels are aligned with the position angle of the X-ray brightness distribution they measured ($\theta = 19\fdg5 \pm 0\fdg6$). However, this map of residual structure was computed by subtracting a radially-symmetric surface brightness profile for the cluster, so a map of excess light in alignment with the cluster's major axis would also be expected by any elongated cluster profile. 

Nevertheless, the fact that different results can be derived from the same observations implies that the X-ray data alone are not able to confirm the presence of filaments. To identify filaments, we also need an observable that can be localized to a three-dimensional position and is also co-spatial with the X-ray gas. Galaxies fulfill this role, but only with sufficient numbers of spectroscopic redshifts to constrain their distance and only with enough observational coverage to identify regions with and without filaments. 

\section{Optical Observations} \label{sec:optical}

\subsection{Spectroscopy}
We conducted a spectroscopic campaign across Abell 133 over four years using the 
Inamori-Magellan Areal Camera and Spectrograph \citep[IMACS,][]{2011PASP..123..288D} on the Magellan Baade telescope. Observations were conducted during four runs: 2011 September/October, 2013 September, 2014 July, and 2014 September. In the course of these runs, we observed 29 different masks, covering 3680 objects (not corrected for duplicates). The details of each mask are given in Table \ref{tab:IMACS}. ${\rm N}_{\rm Obj}$, ${\rm N}_{\rm C}$, and ${\rm N}_{\rm U}$ are the number of slits, galaxies with secure redshifts consistent with Abell 133 (defined for this purpose as galaxies with velocities within $2500\ {\rm km}\ {\rm s}^{-1}$ of the cluster redshift), and the number of spectra for which we were not able to securely obtain redshifts, respectively. For all of the IMACS observations, we used $1\farcs 0$ wide slits and the f/2 (short) camera, operated with the 300 l/mm grating ($1.35$  \AA\ ${\rm pix}^{-1}$, $R \sim 1000$). 

\begin{figure*}
\begin{center}
\includegraphics[width=\textwidth]{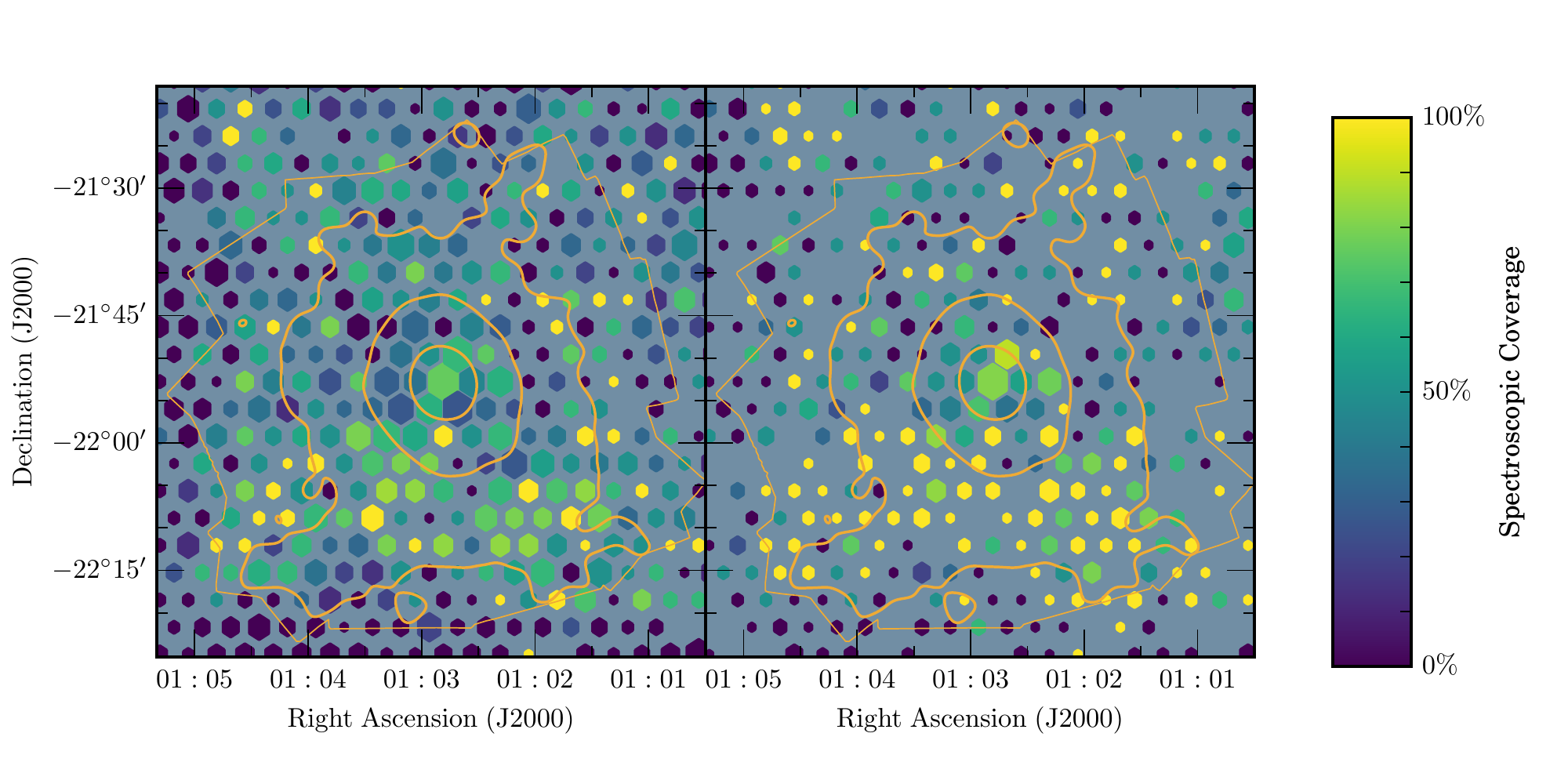}
\end{center}
\caption{Spectroscopic completeness around Abell 133 for all blue ($g-r < 1$) galaxies brighter than $r=21$ mag ({\bf left}) and $r=19.5$ mag ({\bf right}). The former magnitude limit is roughly the extent of our spectroscopic coverage, while the latter is the limit we adopt for much of our analysis. Hexes are colored based on the percentage of galaxies in the hex region that have a spectroscopic redshift, including those from literature. The area of each hex is proportional to the number of galaxies inside its region, with the largest hex containing 17 galaxies. The extent of the X-ray field of view and the contours of the X-ray emission are shown by the thin and thick orange lines, respectively.} \label{fig:spec_coverage_hexes}
\end{figure*}

Targets were selected using archival imaging from the Canada-France-Hawaii Telescope (CFHT) to a limiting magnitude of $r \sim 22$. CFHT $r$-band imaging was taken with MegaCam \citep{2003SPIE.4841...72B} and processed with MegaPipe pipeline \citep{2008PASP..120..212G}. Mask locations were distributed around Abell 133, and the mask design software, {\tt maskgen}\footnote{\href{http://code.obs.carnegiescience.edu/maskgen}{http://code.obs.carnegiescience.edu/maskgen}}, balanced putting as many slits on objects as possible while giving priority to brighter objects. We did not include color information in our selection. We show the distribution of galaxies that we were able to successfully obtain spectroscopic redshifts for in Figure \ref{fig:spec_coverage_hexes}. 

IMACS data were reduced following standard practices. The data reduction pipeline will be described in a future paper (Kelson et al. in prep). It is a semi-automated set of routines for identifying and isolating individual spectra in multi-slit observations using Coherent Point Drift \citep{2009arXiv0905.2635M} to bootstrap global wavelength mappings across the full field of IMACS. Once these mappings are derived, the slitlets are processed using routines described in \citet{2000ApJ...531..159K} and \citet{2003PASP..115..688K}. Flat-fielding was performed with quartz-halogen lamp images taken immediately after our science images; wavelength calibration was achieved using HeNeAr spectral lamp exposures taken after the flat field images. After being individually processed, multiple exposures of the same slit were combined. Background subtraction was performed using B-splines \citep{2003PASP..115..688K}; this technique identifies and masks out cosmic rays during the combining of individual 2D spectra, thus obviating the need for any cosmic ray cleaning. 

\subsection{Fitting Redshifts}
For each extracted slit in each of the masks, we attempted to measure redshifts using cross-correlation techniques\footnote{Described at this link: \url{http://code.obs.carnegiescience.edu/Algorithms/realcc}}. Spectral extraction and redshift template fitting were performed using the tools developed by \citet{2018AAS...23114942L}. We used the spectral templates from the Sloan Digital Sky Survey \citep[SDSS,][]{2004AJ....128..502A} as comparisons for our cross-correlation. For most objects, we only used the galaxy templates (23--28), but, for a number of objects we had identified by eye as being broad-line galaxies, we also used templates 30 and 33. We first performed two automatic passes with the redshift finder, flagging galaxies with apparent emission lines between the two passes. After visually inspecting all of these results, we then manually ran the redshift finding routine on every galaxy without a definitively clear redshift. We marked any galaxy whose redshift we could not conclusively determine as ``unsure.'' Those galaxies whose spectra had significant issues that made measuring a redshift impossible were marked as ``bad.'' We also identified any stars that had been observed, primarily through identification of narrow absorption at H$\alpha$.
\begin{figure*}
\begin{center}
\includegraphics{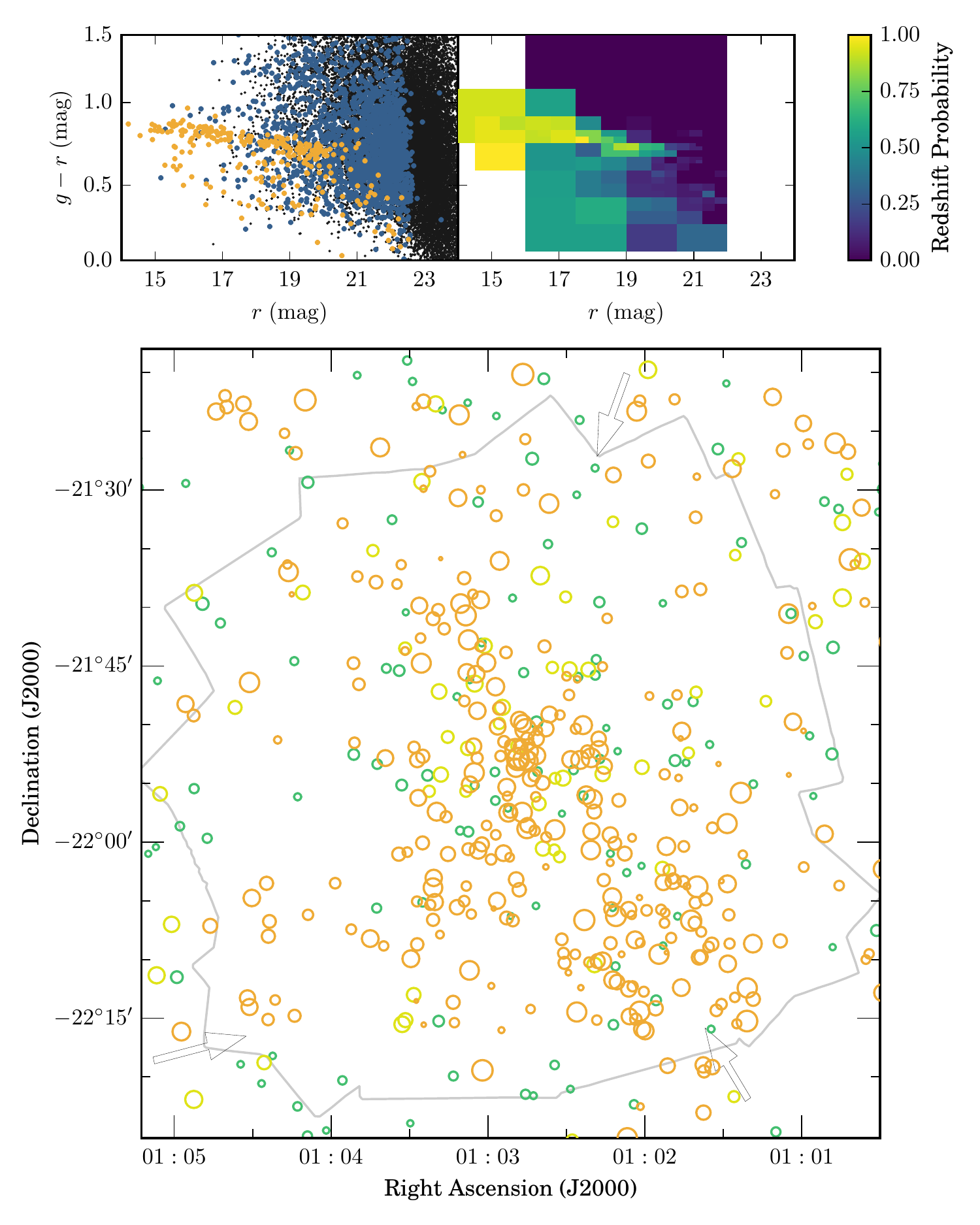}
\end{center}
\caption{Photometric distribution of galaxies around Abell 133. {\bf Top}: $g-r$ vs. $r$ color-magnitude diagram for all galaxies within half a degree of the cluster center. {\bf Top left}: photometry of individual galaxies, where orange points are cluster members ($|v - v_c| < 2500\ {\rm km}\ {\rm s}^{-1}$), blue are spectroscopic non-members, and black are galaxies without spectra. {\bf Top right}: fractional distribution of cluster members in dynamically sized bins (as described in Section \ref{ssec:merging_cats}). {\bf Bottom}: distribution of potential cluster members. Spectroscopic members are shown in orange, while galaxies in bins with probabilities of at least 75\% (50\%) are shown in yellow (green). Points are sized proportional to their brightness. The extents of the X-ray observations are shown in gray, and the arrows indicating the location of the filaments from Figure \ref{fig:two_xray_images} are included. Our spectroscopic incompleteness is not causing us to miss any important structures.} \label{fig:cmd_spec}
\end{figure*}

In total, we measured reliable spectroscopic redshifts for 2878 galaxies. To verify the accuracy of these measurements, we compared multiple observations of the same object. For 75 unique pairs of different observations of the same target, the accuracy -- as characterized by $\sigma = 1.48 \times \tilde{|d|}$, where $\tilde{|d|}$ is the median absolute deviation -- was $\sigma = 0.00020$. We found no change in this value by only considering either galaxies with or without emission lines. This accuracy is within the level needed for our science, corresponding to a velocity error of $\sigma_v \approx 60\ {\rm km}\ {\rm s}^{-1}$. 

\subsection{Photometry}
While the target selection for spectroscopy was performed using CFHT imaging data, deeper $griz$Y photometry of a much larger area became publicly available with the Dark Energy Survey (DES) Data Release 1 \citep{2018arXiv180103181A}. The details of the photometric processing in this database are given by \citet{2018PASP..130g4501M}. Our photometric catalog includes every object within a $3^\circ \times 3^\circ$ box centered on Abell 133 with ${\tt FLAGS} < 4$ in all five bands. After integrating our spectroscopic redshifts into the photometric data, we removed stars from our catalog using the {\tt EXTENDED\_COADD} parameters suggested by \citet{2018arXiv180103181A}, excising objects with ${\tt EXTENDED\_COADD} < 2$ (high-confidence and likely stars). This selection caught 98.4\% of known stars, while it only would have excised 1.3\% of spectroscopically-confirmed galaxies. The spectroscopic coverage of the photometric sample within $0\fdg5$ of the cluster center is shown in Figure \ref{fig:cmd_spec}.

\subsection{Merging Catalogs}
\label{ssec:merging_cats}
To unite our photometric and spectroscopic catalogs, we first checked to see if each photometric object had a match in the spectroscopic database. For galaxies with multiple spectroscopic matches, we excluded any spectra marked ``unsure'' or ``bad.'' If the object had multiple good observations, we used the redshift information from the spectrum with the best signal-to-noise ratio. Photometric objects with no good spectroscopic match were then checked against several supplementary catalogs. First, we checked to see if these objects were included in the fourth United States Naval Observatory (USNO) CCD Astrograph Catalog \citep[UCAC4,][]{2013AJ....145...44Z}; we classified everything with a match in the UCAC4 with object classification flag and g-flag values of 0 as a star (these two flags selected objects described as a ``good, clean star'' with no obvious galaxy match). If the object was still unidentified, we then checked for matches in the NOAO Fundamental Plane Survey \citep[NFPS,][]{2004AJ....128.1558S}, the 6dF Galaxy Survey \citep{2009MNRAS.399..683J}, the Southern Sky Redshift Survey \citep[SSRS,][]{1998AJ....116....1D}, two AGN redshifts from \citet{1998MNRAS.299.1047W} and \citet{1999ApJS..122...29B}, and a spectroscopic survey of Abell 133 by \citet{1997astro.ph..9036W}. The final counts of spectra from each survey are given in Table \ref{tab:SpecSum}, as are the number of galaxies used in individual steps of the following analysis.

\begin{deluxetable}{lr}
\tablecaption{Summary of Spectroscopic Observations}
\label{tab:SpecSum}
\tablewidth{0pt}
\tablehead{
\colhead{Category} & \colhead{Number} }
\startdata
Slits Observed & 3680 \\
Redshifts Obtained & 2878 \\
NFPS \citep{2004AJ....128.1558S} & 29\tablenotemark{a}\\
6dF Galaxy Survey \citep{2009MNRAS.399..683J} & 72\tablenotemark{a} \\
SSRS \citep{1998AJ....116....1D} & 0\tablenotemark{a}\\
\citet{1997astro.ph..9036W} & 60\tablenotemark{a}\\ 
$|\Delta v| < 2500\ {\rm km}\ {\rm s}^{-1}$, $r \leq 1500\arcsec$ & 242\\ 
$|\Delta v| < 2500\ {\rm km}\ {\rm s}^{-1}$, $r \leq 2000\arcsec$ & 302\\ 
In Potential SW Substructure & 71\\
In Potenntial NE Substructure & 24\\
Used to Calculate $z_c$, $\sigma_v$& 254\\ 
In annular region used for Cross-Correlation& 132\\ 
\enddata
\tablenotetext{a}{Additional redshift values used}
\end{deluxetable}

While we discuss the completeness of our spectroscopic observations in Appendix \ref{sec:spec_comp}, we briefly discuss the key points here. For galaxies within $1500\arcsec$ of the cluster center, we reach a cumulative 50\% completion fraction at $r \sim 19.4$, and the faintest galaxies with redshifts are at $r \sim 22.2$. We have a slight bias in the angular distribution of galaxies with redshifts, such that the number of galaxies to the south is larger than the number to the north, again within $1500\arcsec$. This deviation is about twice the deviation expected from Poissonian sampling. However, due to the available imaging data at the time we made our masks, our spectroscopic coverage extends north of the X-ray field of view but does not extend all the way to the field's southern border. We show in Figure \ref{fig:spec_coverage_hexes} the spatial coverage of our spectroscopic sample. This distribution should be considered when we discuss the distribution of cluster members.

The distribution of galaxies with velocities near that of the cluster is shown in the bottom panel of Figure \ref{fig:cmd_spec}. This is supplemented by the positions of galaxies without spectroscopic redshifts but with $g-r$ colors and $r$ magnitudes consistent with cluster galaxies. To determine this, we slice the color-magnitude plane into a $6 \times 6$ grid covering $-0.3 < g-r \leq 1.5$ and $13 < r \leq 22$; each bin is assigned a probability based on the fraction of galaxies with spectroscopic redshifts that are within $\Delta v \leq 2500\ {\rm km}\ {\rm s}^{-1}$ of the cluster redshift, with a minimum of 16 total galaxies with measured redshifts per bin. We then bisect each bin in color and magnitude space, making a $12 \times 12$ grid, and repeat the probability measurement, although now only requiring 8 galaxies per bin. This continues until we have a $192 \times 192$ grid covering the photometric area, requiring 8 galaxies per bin for all subdivisions.

Galaxies without a spectroscopic redshift are then assigned a cluster probability based on the value of the smallest bin they occupy. The values of these bins are shown in the upper-right panel of Figure \ref{fig:cmd_spec}. As is shown in the bottom panel of that figure, there is a missing population of galaxies without spectroscopic redshifts that may be cluster members, but {\it there is no physical structure that is not being identified by the redshifts.} We see in Figure \ref{fig:cmd_spec} that the photometric distribution of galaxies has the same morphology as the X-ray emitting gas seen in Figure \ref{fig:two_xray_images}  -- there is extended structure to the SE, the SSW, and to the N. We discuss the galaxy distribution in more detail in Section \ref{sec:optical_ids} .

\begin{figure*}
\begin{center}
\includegraphics[width=\textwidth]{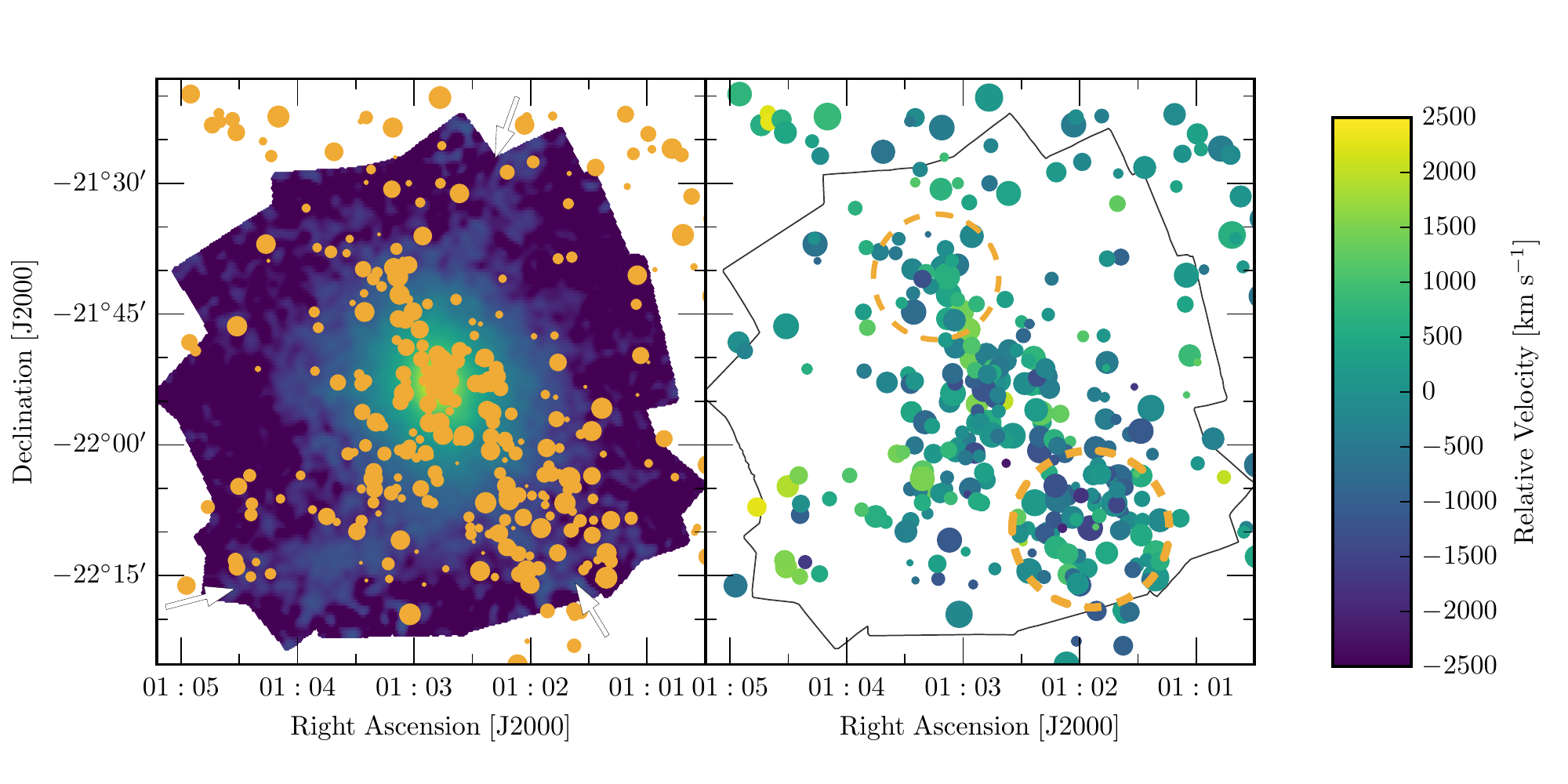}
\end{center}
\caption{The spatial distribution of galaxies around Abell 133. {\bf Left:} Map of spectroscopic members with reference to the X-ray emission. Positions of every galaxy with a redshift close to that of the cluster ($0.047 \leq z \leq 0.065$) are marked with orange points. Radii of these points scale linearly with the $r$ magnitude of the galaxy (brighter galaxies are larger). The X-ray emission and arrows are the same as is shown in the right panel of Figure \ref{fig:two_xray_images}. {\bf Right:} Velocity map of the galaxies shown on the left panel. Velocities are computed from spectroscopic redshifts, as described in the text. Two regions of potential structure are indicated by dashed orange circles; neither shows any statistical evidence of being different from the velocity profile of the rest of the cluster. The colorbar only applies to the right panel.} \label{fig:specs_on_x}
\end{figure*}

\section{The Galaxies Around Abell 133}\label{sec:galaxies}
\subsection{Velocity Structure Around Abell 133}\label{sec:VelocityStructure}
With our spectroscopic sample, we were able to investigate the galaxy velocity distribution of Abell 133 and, in particular, to look for possible substructure in the galaxy distribution that corresponded with the observed X-ray structures. To that end, we first considered the distribution of cluster galaxy velocities. For galaxies located within $1500 \arcsec$ of the cluster center, we first compute a cluster redshift, $z_c$, using the biweight estimator of \citet{1990AJ....100...32B}. We calculate the values of $z_c$ and the velocity dispersion, $\sigma_v$, for all galaxies with redshifts $0.0408 \leq z \leq 0.0708$. By converting a redshift deviation to a velocity deviation using Equation \ref{eqn:vel_calc}, we find that $z_c = 0.055865$ and $\sigma_v = 813 \pm 44 \ {\rm km}\ {\rm s}^{-1}$. Using the $\sigma_v - T_X$ scaling relation presented by \citet{2016MNRAS.463..413W}, the expected nominal $\sigma_v$ for gas of this temperature \citep[3.61 keV,][]{2006ApJ...640..691V} is $\sigma_v = 791 \ {\rm km}\ {\rm s}^{-1}$, which is a $1\sigma$ agreement with our reported value.

Previous work by \citet{1997astro.ph..9036W} identified an offset between the cluster velocity and the velocity of the cD galaxy, as parametrized by the Z-score \citep{1991ApJ...383...72G}. Following the standard convention \citep[e.g.,][]{1994ApJ...422..480B} that a velocity offset is significant when the 90\% confidence intervals of the Z-score do not bracket 0, their result of ${\rm Z}_{\rm score} = 0.260^{+0.166}_{-0.164}$ (90\% confidence intervals) were taken as evidence for a dynamically young cluster. \citet{2010ApJ...722..825R} built on that result, finding ${\rm Z}_{\rm score} = 0.256^{+0.170}_{-0.162}$. However, these analyses were built on an assumed cD velocity of $v_{cD} = 17051\ {\rm km}\ {\rm s}^{-1}$ \citep{1997astro.ph..9036W}; if we use the measured velocity from the NFPS \citep{2004AJ....128.1558S}, $v_{cD} = 16783\ {\rm km}\ {\rm s}^{-1}$, and the new redshifts presented here, then the Z-score is ${\rm Z}_{\rm score} = 0.041^{+0.111}_{-0.106}$. It is not clear why the two measurements of the velocity differ by almost $300\ {\rm km}\ {\rm s}^{-1}$, and neither \citet{2009MNRAS.399..683J} nor we obtained an independent redshift of the cD galaxy. However, \citet{2004AJ....128.1558S} compared the NFPS redshifts to literature values for 3004 other galaxies, and found a median offset of $+6 \pm 2\ {\rm km}\ {\rm s}$. 
We found redshifts for 16 galaxies also in the NFPS sample, and the median offset (assuming $v = cz$) was $-49 \pm 72\ {\rm km}\ {\rm s}^{-1}$. Assuming then that the NFPS values are accurate, the cD velocity is not unusual.

We show the distribution of galaxies with velocities within 2500 ${\rm km}\ {\rm s}^{-1}$ of the cluster in the left panel of Figure \ref{fig:specs_on_x} (as shown below, this includes the entire cluster population). 
On the right panel of Figure \ref{fig:specs_on_x}, we show the velocity structure of these galaxies. In particular, we highlight two regions of enhanced density, which are marked by dashed orange circles in the figure. 
These two regions correspond to density enhancements highlighted by \citet{2006A&A...450....9F} in their analysis of archival photographic plate images.
\begin{figure*}
\includegraphics{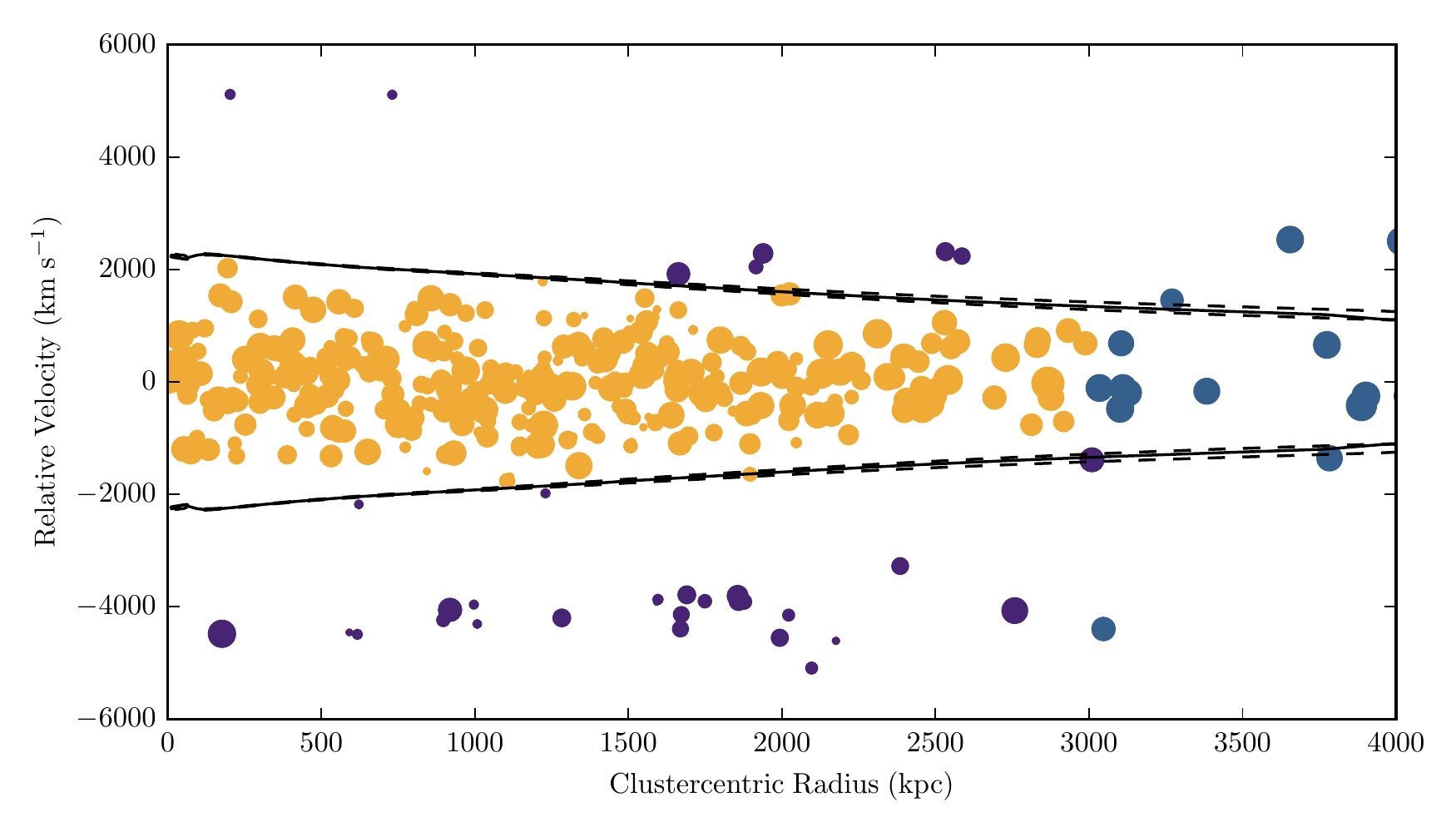}
\caption{The measured velocity structure around Abell 133 for all galaxies with known redshifts. A fit to the line-of-sight escape velocity is shown with solid lines, while uncertainties in this velocity are marked by the dashed lines. Galaxies contained within these velocity points are plotted in orange, galaxies with velocities greater than the cluster escape velocity are shown in purple, and galaxies beyond $2 \times R_{200}$ are marked in blue. The escape velocity was calculated from an NFW fit and the mass model of this cluster from \citet{2006ApJ...640..691V}, as described in the text. Points are sized based on their brightness, with brighter galaxies being larger on this plot. We see no significant substructure in this plot, implying that there is no kinematic disruption.} \label{fig:caustic_curve}
\end{figure*}

To determine if these regions are kinematically different from the rest of the cluster, we perform a two-sample Kolmogorov-Smirnoff test. We compare the galaxies inside those projected regions with the rest of the galaxies within 1500 kpc of the cluster center, with the null hypothesis that the two samples are drawn from the same kinematic distribution. We find no statistical evidence to disprove the null hypothesis ($p = 0.72$ and $p = 0.28$ for the NE and SW clumps, respectively), meaning that these clumps are only spatial overabundances and not meaningful kinematic substructures. We note that the SW clump was previously identified by \citet{1999A&A...351..883K} as being evidence of subclustering in Abell 133, while \citet{2013MsT..........GJ} identified both clumps in a map of red sequence galaxies.

To constrain the presence of substructure in the cluster, we considered the velocity space around Abell 133's position and redshift. In Figure \ref{fig:caustic_curve} we show the velocity relative to the cluster as a function of the clustercentric radius for galaxies with spectroscopic redshifts. While a number of works \citep{1997ApJ...481..633D,1999MNRAS.309..610D,2011MNRAS.412..800S} have used the velocity-radius distribution to derive contours of escape velocity and, from that, a mass profile, we have a relatively low number of confirmed members \citep[see][]{2017ApJ...834..204G} with a non-uniform selection function. Therefore, we do not attempt to fit the cluster's profile and instead use previous fits to Abell 133's mass distribution to derive the constraints on escape velocity. Assuming a density profile of \citet[hereafter NFW]{1996ApJ...462..563N}, the gravitational potential can be characterized as
\begin{equation}\label{eqn:phi_r}
\Phi (r) = \frac{-G M(r)}{r} - 4 \pi {\rm G} \rho_0\,  \frac{R_S^2}{1 + r/R_S}.
\end{equation}
Here, G is the gravitational constant, $R_S$ is a scale radius of the NFW profile, and $\rho_0$ is a characterization of the NFW profile, defined as four times the density at $r = R_S$. \citet{2006ApJ...640..691V} reported a value of $R_S$ for this cluster, $R_S = 1239.9\ {\rm kpc}$, while the value of $\rho_0$ can be calculated from their reported values of $R_{500}$ and $M_{500}$: $\rho_0 = 6.09 \times 10^{-27}\ {\rm g}\ {\rm cm}^{-3}$. Assuming Newtonian dynamics and spherical symmetry, the escape velocity is related to the gravitational potential through
\begin{equation}
v_{esc}^2 = -2 \Phi (r).
\end{equation}
We make the simplifying assumption the escape velocity in the line-of-sight (which is what is measured by redshifts) is 
\begin{equation}
v_{esc,los}^2 = \frac{1}{3} v_{esc}^2,
\end{equation}
which is to say that the velocity distribution is assumed to be isotropic. To calculate $\Phi (r)$ using Equation \ref{eqn:phi_r}, we use the mass profile of Abell 133 from \citet{2006ApJ...640..691V}. Upper and lower errors on $v_{esc,los}$ are based on the errors in $M(r)$.

In Figure \ref{fig:caustic_curve}, we show that the structure around the cluster is mostly consistent with being bound to Abell 133. The only exception is a small number of galaxies around $-4500\ {\rm km}\ {\rm s}^{-1}$ from the cluster. This corresponds to a redshift of $z \sim 0.04$, which is ${\sim}65\ {\rm Mpc}$ in the foreground, assuming the velocity offset is due to the Hubble expansion. We do not see any other major structures in phase-space out to relative velocities of at least 10,000 ${\rm km}\ {\rm s}^{-1}$. Were the X-ray emissions indicative of ongoing merging, we would expect to see structure in phase-space \citep[e.g.,][]{2016ApJS..224...33B}. 

From a kinematic perspective, then, Abell 133 is not particularly disturbed. The cD galaxy is not significantly offset from the cluster's rest velocity, the observed galaxy overdensities are not kinematically distinct from the cluster at large, and there is no obvious substructure in velocity space. Arguments for merger activity based on X-ray analysis of the cluster in \citet{2010ApJ...722..825R} are based on assumptions about the cD's status that are not supported by our analysis, as well as a marginal detection of a hot spot in the X-ray gas seen by {\it Chandra} but not seen by {\it XMM-Newton}. The diffuse radio structure identified by \citet{2001AJ....122.1172S} as a radio relic and the presence of a tongue in the X-ray image are consistent with a buoyantly rising bubble \citep{2002ApJ...575..764F,2004ApJ...616..157F}. The detection of an H$\alpha$ filament by \citet{2010ApJ...721.1262M} and the low central entropy \citep{2009ApJS..182...12C} further support the notion that this radio emission is associated with the cooling flow and not a merger shock \citep[e.g.,][and references within]{2018ApJ...858...45M}.


\subsection{Optical Counterparts of X-ray Structure}
\label{sec:optical_ids}

In addition to analyzing the kinematics of Abell 133, our spectroscopic sample allows us to investigate -- both qualitatively and quantitatively -- how the galaxy distribution compares to that of the extended X-ray emission. \citet{2013HEAD...1340101V} proposed that three filaments are visible in the X-ray image, which are shown in Figure \ref{fig:two_xray_images}. The galaxy distribution also has a triaxial distribution, with extensions to the southwest, southeast, and north, which can be seen in Figure \ref{fig:specs_on_x}. Here, we assess how closely the galaxy distribution relates to the filaments identified in X-rays.

\begin{figure*}
\begin{center}
\includegraphics{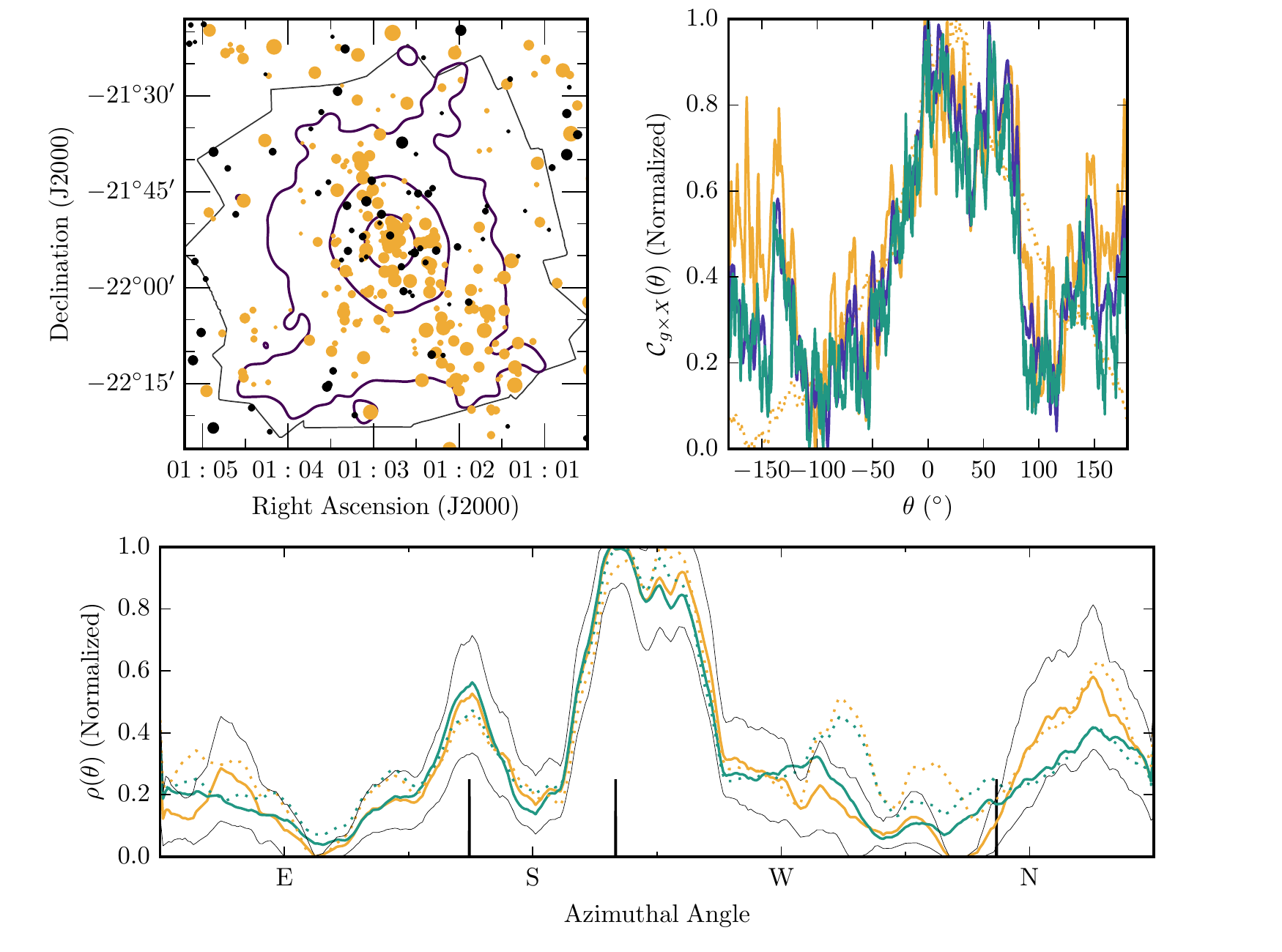}
\end{center}
\caption{{\bf Top left}: distribution of spectroscopic (orange) and photometric (black) cluster members brighter than $r < 19.5$ mag. Spectroscopic members are those galaxies with redshifts within $5\sigma$ of the cluster redshift, while photometric members are those galaxies with a redshift probability of $p > 0.5$, as defined in Section \ref{ssec:merging_cats} and Figure \ref{fig:cmd_spec}. The X-ray field of view is marked by gray lines, while the smoothed contours of the X-ray emission are shown in purple. The triaxial structure of the galaxy distribution matches that of the X-ray gas. {\bf Top right}: cross-correlation between the galaxy positions and the X-ray emission. Positive $\theta$ corresponds to a clockwise rotation of the galaxy population. The cross-correlation between all spectroscopic members (the orange galaxies on the left panel) is marked by the dashed orange line; the same population but only for those galaxies at $r > R_{500}$ is shown with the solid orange line. In purple, we show the cross-correlation between all galaxies on the top left plot outside $R_{500}$, and in green we show the cross-correlation between all galaxies shown in the top left panel outside $R_{500}$ against a binary X-ray map. In all cases, the signal peaks around $0^\circ$, implying that there is a statistical match between the X-ray structures identified by \citet{2013HEAD...1340101V} and the galaxy positions. {\bf Bottom}: Azimuthal density profile of the galaxy distribution when adjusted for inverse completeness (orange) or using photometric probabilities (green). The angular density inside $750\arcsec - 1500\arcsec$ is shown in solid lines and inside $750\arcsec - 1750\arcsec$ with dashed lines. $1\sigma$ confidence intervals of the inverse completeness weighted galaxies in the $750\arcsec - 1500\arcsec$ annulus, as determined through bootstrap resampling, are shown with gray lines. The three filaments identified by \citet{2013HEAD...1340101V} are marked with vertical black lines. The two prominent peaks to the SSE and SW are matched in both distributions, while both distributions also show excess to the N. As with the qualitative analysis, we cannot confirm the NW X-ray filament through the azimuthal galaxy density distribution.} \label{fig:xray_crosscor}
\end{figure*}

The most prominent component of the galaxy distribution is the extension to the southwest. In Figures \ref{fig:cmd_spec} and \ref{fig:specs_on_x} we can see that the distribution of galaxies extends beyond the X-ray field of view. These galaxies lie along the X-ray filament identified by \citet{2013HEAD...1340101V} as well as the single filament identified by \citet{2014MNRAS.437.1909M}. However, the peak of X-ray intensity shown by \citet{2014MNRAS.437.1909M} is slightly to the west of that proposed by \citet{2013HEAD...1340101V}; the \citet{2013HEAD...1340101V} measurement appears to be better matched by the galaxy distribution.


In the X-ray map of \citet{2013HEAD...1340101V}, the most prominent filament is to the southeast; this filament is the target of future {\it Chandra} observations \citep{2017cxo..prop.5264K}. In the distribution of galaxies shown in Figure \ref{fig:specs_on_x}, however, the excess of galaxies appears slightly north of the X-ray structure. While this may be a real offset, there is a compelling case that this offset is an imprint of our spectroscopic campaign. As discussed in Appendix \ref{sec:spec_comp} and shown in Figure \ref{fig:spec_coverage_hexes}, the input catalog used for target selection did not extend as far south as the X-ray image. Additionally, IMACS masks are circular, so that the bottom edges of the field will be undersampled. In Figure \ref{fig:cmd_spec}, we identified four likely cluster members members in the X-ray filament that were not observed; furthermore, a number of galaxies may have been missed with spectroscopy and been too faint for our probabilistic model to identify, but are within the brightness limits of this survey. As the galaxy distribution extends out toward this X-ray filament, our qualitative analysis suggests that this filament is also matched by the galaxy distribution.

Finally, we consider the structure to the north. In contrast to the southeastern filament, here our spectroscopic coverage extends well to the north of the northern border of the X-ray field of view. Considering only the distribution of spectroscopically confirmed cluster galaxies (shown in the left panel of Figure \ref{fig:specs_on_x}) without regard to their distribution of brightness, a linear feature can be seen extending to the northeast from the cluster center in the direction highlighted by \citet{2014MNRAS.437.1909M}. There are few galaxies on the northern filament identified by \citet{2013HEAD...1340101V}, although we note that we identified several galaxies in Figure \ref{fig:cmd_spec} that are along this filament and have the photometry of cluster members. In the top left panel of Figure \ref{fig:xray_crosscor}, we show spectroscopic members and probable members brighter than $r = 19.5$ (our 50\% completeness limit, as discussed in Appendix \ref{sec:spec_comp}). In this view, the northern galaxy distribution bends to the northwest, slightly above the filament detected by \citet{2013HEAD...1340101V}. From the optical data alone we can see that there is some structure to the north, but it is difficult to definitively constrain the path of that structure. 

To provide a quantitative measure of how well the galaxy population aligns with the X-ray emission, we consider a cross-correlation between the X-ray flux and galaxy positions. If the galaxy distribution maps out the underlying distribution of X-ray emitting gas, we would expect that the cross-correlation would be peaked in the current alignment and would decrease as we move the galaxies. In particular, by rotating the galaxies around the cluster's center, the amplitude of the cross-correlation should decrease (although a small bounce-back is expected as filaments in the galaxy distribution rotate onto the next X-ray filament). To test this prediction, we consider this formula for the galaxy X-ray cross-correlation:
\begin{equation}
{\cal C}_{g \times X} (\theta) = \frac{\sum\limits_{i}^n X_{x_i,y_i}(\theta) \times W_i}{\sum\limits_{i}^n W_i} .
\end{equation}
Here, $X_{x_i,y_i}(\theta)$ is the X-ray flux at the position of galaxy $i$ when rotated around the cluster center by angle $\theta$, $W_i$ is the weight applied to that galaxy, $n$ is the number of galaxies being considered, and $\theta$ is defined such that a rotation of $\theta > 0$ is clockwise. In this equation we discard any galaxy that is not in the X-ray field of view at angle $\theta$. $1\sigma$ confidence intervals are found by bootstrap resampling. 

We first consider all galaxies with spectroscopic velocities within $2500\ {\rm km}\ {\rm s}^{-1}$ of the cluster, setting all of their weights to the inverse completeness, $W_i = g_i^{-1}$, where $g_i$ is the fraction of galaxies within $5\arcmin$ of galaxy $i$ brighter than the magnitude cut that have spectroscopic redshifts. ${\cal C}_{g \times X} (\theta)$ is maximized at $\theta = 21.0^{+ 17.0}_{- 20.0} {}^\circ$. The values of ${\cal C}_{g \times X} (\theta)$, normalized from 0 to 1, are shown in Figure \ref{fig:xray_crosscor}. As this is dominated by the signal in the center of the cluster, where the X-ray emission is strongest, we then exclude all galaxies within $1007\arcsec$ of the cluster center \citep[this is $R_{500}$, as reported by ][]{2006ApJ...640..691V}; with this change, the cross-correlation peaks at $\theta = -3.0^{+ 58.0}_{- 0.5} {}^\circ$. Here, the cross-correlation is bi-modal; the large number of galaxies observed on the SW filament rotate onto the strong X-ray emission of the SE filament around $\theta \sim 55^\circ$, boosting the signal. 

To account for our spectroscopic incompleteness, we also include galaxies without spectroscopic redshifts but that are potentially cluster members. Weighting these galaxies with $W_i = f(g,r)$, where $f(g,r)$ is the probability of being a cluster member based on the galaxy's $g$ and $r$ magnitudes as discussed in Section \ref{ssec:merging_cats}, and only considering galaxies with $f(g,r) \geq 0.5$, the cross-correlation peaks at $\theta = 0.0^{+ 55.0}_{- 4.0} {}^\circ$. Finally, to account for any bias in the X-ray flux, we construct a binary map of X-ray flux, where $X=1$ where $f \gtrsim 4.3 \times 10^{-8}\ {\rm count}\ {\rm s}^{-1}\ {\rm arcsec}^{-2}$ and $X=0$ otherwise. In this case, the maximum cross-correlation is at $\theta = 0.0^{+ 55.0}_{- 28.0} {}^\circ$. 

As the distribution is bimodal, we also calculate 68\% confidence intervals when discarding bootstrapped values between $45^\circ$ and $85^\circ$. For the four scenarios we tested, the best fit values (with percentage of values discarded in parentheses) are $\theta = 21.0^{+ 13.5}_{- 20.0} {}^\circ$ (all galaxies, 5.5\% excluded), $\theta = -3.0^{+ 35.5}_{- 25.0} {}^\circ$ (outer galaxies, 28.6\% excluded), $\theta = 0.0^{+ 10.0}_{- 4.5} {}^\circ$ (outer photometric galaxies, 35.7\% excluded), and $\theta = 0.0^{+ 13.0}_{- 4.5} {}^\circ$ (outer photometric galaxies with binary X-ray map, 32.7\% excluded). Using the mean and standard deviation of ${\cal C}_{g \times X} (\theta)$ between $\theta = 100^\circ$ and $\theta = -50^\circ$ as a proxy for the noise, the peaks in ${\cal C}_{g \times X} (\theta)$ correspond to ${\sim}4 - 5\sigma$ detections. As an alternative measure of the detection significance, we computed an expectation value of ${\cal C}_{g \times X}$ for the four cases, assuming the same X-ray map but with bootstrapped populations of galaxies. This bootstrapping set each galaxy to a random angle, but kept the clustercentric radii of the parent galaxy samples. Here, the peak of ${\cal C}_{g \times X}$ seen in the four cases has significances of $2.1\sigma$ (which is expected with most of the X-ray emission being in the cluster center), $3.9\sigma$, $5.6\sigma$, and $5.6\sigma$.

\begin{figure*}
\begin{center}
\includegraphics[width=\textwidth]{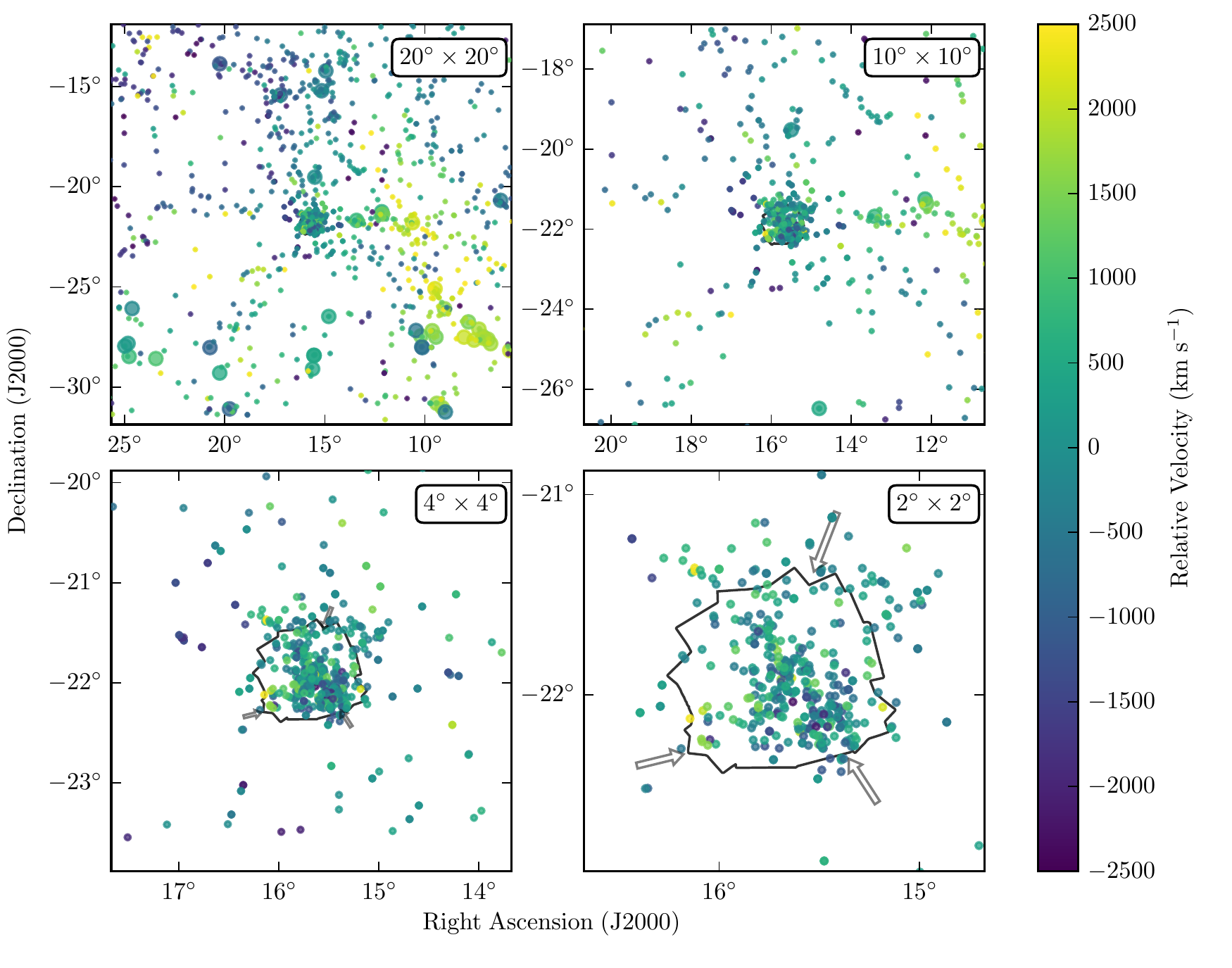}
\end{center}
\caption{The large-scale structure around Abell 133. The four panels are centered on the cluster, but cover angular boxes with sizes of $20^\circ$, $10^\circ$, $4^\circ$, and $2^\circ$ from left to right then top to bottom. Galaxies, groups, and clusters are all shown; groups and clusters are marked with larger points. Points are colored based on their velocity relative to Abell 133. We note that, assuming that motions are due solely to Hubble expansion, the depth of these images is ${\sim}70\ {\rm Mpc}$, which is also roughly the size of the largest box. The outer edges of the X-ray images are marked with dark gray lines, and the directions of the filaments reported from the X-ray image are marked by arrows. Cosmic filaments are apparent in the larger fields of view, but within $2^\circ$ of the cluster the filaments become difficult to trace due to the relatively low surface density of points.} \label{fig:full_zoomout}
\end{figure*}

Our cross-correlation analysis shows that there is a statistically significant alignment between the galaxies and the X-ray distribution. However, to characterize how well each individual filament maps to the positions of galaxies, we also measured the angular distribution of the galaxy population. To do this, we measured an azimuthal galaxy density profile
\begin{equation}\label{eqn:galaxy_power}
\rho(\theta) = \sum_{i \in A} W_i,
\end{equation}
where $W_i$ is the weight assigned to each galaxy and $A$ is the set of all weights within $2.5^\circ$ of $\theta$. To account for the finite number of samples available, each point is smoothed by a Gaussian kernel of width $\sigma = 9^\circ$. Our galaxy sample is limited to those galaxies brighter than $r=19.5$ mag, and completeness is computed only for galaxies bluer than $g-r = 1.0$. $W_i$ is calculated in two ways: inverse completeness and photometric probability. Here, the inverse completeness is based on the total fraction of galaxies within $2.5^\circ$ of a given galaxy within the magnitude, color, and radial ranges that have spectroscopic redshifts.
 
We measure $\rho(\theta)$ in two radial ranges: $750\arcsec - 1500\arcsec$ (the inner boundaries of the X-ray image) and $750\arcsec - 1750\arcsec$ (the approximate edge of most of the X-ray image). These distributions are shown in the bottom panel of Figure \ref{fig:xray_crosscor}. We also show the $1\sigma$ confidence intervals for the former radial range in gray, as determined through bootstrap resampling. 
The galaxy distribution has peaks to the SSE and SW, confirming the presence of those X-ray features. In addition, there is a third peak in the galaxy distribution, which is to the NNE; this is not coincident with a single X-ray structure, but it does align with the broad X-ray impression of structure to the N. As with our qualitative analysis, we cannot confirm the exact path of the northern filament, but we do show that there is an excess of structure in that direction. 

\section{Insights from the Cosmic Web}\label{sec:CosmicWeb}

The distribution of galaxies shown in Figures \ref{fig:cmd_spec}, \ref{fig:specs_on_x}, and \ref{fig:xray_crosscor} 
has a similar morphology to the observed X-ray emission, with structures extending from the cluster to the N, SW, and SSE -- the directions of the three filaments identified by \citet{2013HEAD...1340101V}. However, to confirm the X-ray emission is associated with filaments from the cosmic web, we need to identify the cosmic web filaments that these structures in the galaxy distribution may be connected to. Using the additional spectroscopic redshifts outside the IMACS survey area, we can identify the large-scale filaments that feed into the cluster. In particular, we rely upon the 6dF catalogs \citep{2009MNRAS.399..683J}. 

To better trace the structure of the Cosmic Web, we also considered the positions of groups and clusters of galaxies. Using the NASA/IPAC Extragalactic Database (NED), we assembled a catalog of groups and clusters of redshift $0.04 < z < 0.07$ within a $20^\circ \times 20^\circ$ box centered on Abell 133. Although we used NED to assemble the initial catalog, we trimmed this catalog and confirmed the redshifts using the individual references cited by NED. For Abell clusters, we used the compilation of \citet[and references within]{1999ApJS..125...35S}; other cluster redshifts came from \citet{1994MNRAS.269..151D}, \citet{1996A&A...310...31M}, \citet{1997MNRAS.289..263D}, \citet{1999MNRAS.305..259W}, and \citet{2002MNRAS.329...87D}. We also included groups, whose redshifts are compiled from \citet{2002MNRAS.335..216M}, \citet{2002ApJS..140..239C}, \citet{2013ApJ...776...71C}, and \citet{2015A&A...578A..61D}. We note that, while the Abell catalog is mostly uniform across the sky and that many of our other sources only include one or two clusters or groups, most of our groups come from \citet{2002MNRAS.335..216M} and \citet{2013ApJ...776...71C}, which only cover declinations of $\delta_{2000} \lesssim -27^\circ$. 


\begin{figure*}
\begin{center}
\includegraphics[width=\textwidth]{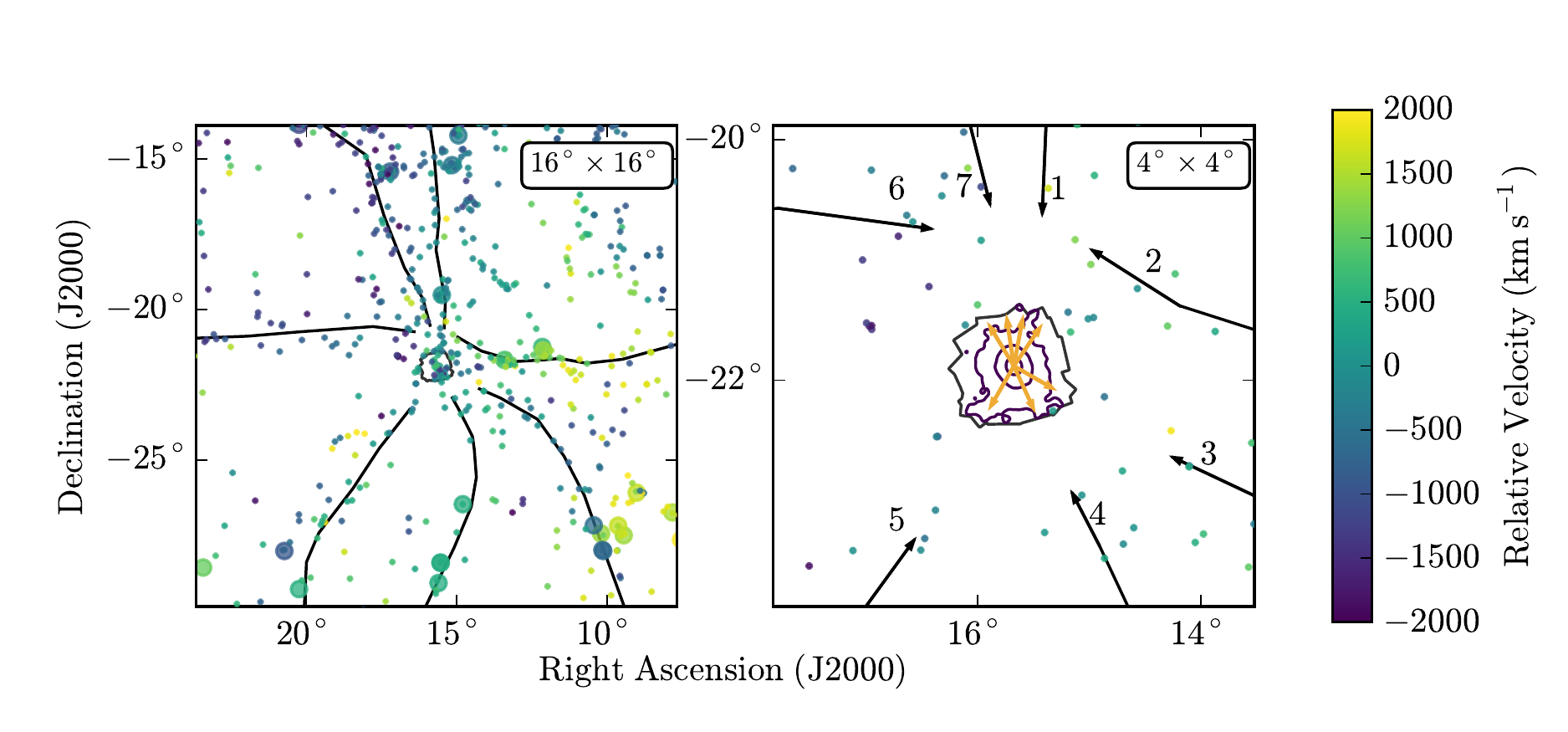}
\end{center}
\caption{
Identified filamentary structure around Abell 133. {\bf Left}: a $16^\circ \times 16^\circ$ square and galaxies within $2000\ {\rm km} {\rm s}^{-1}$ of the cluster redshift; this is effectively a cube of side length 60 Mpc. Filaments are marked with black lines. Due to the paucity of spectroscopic redshifts near the cluster, these lines terminate before connecting to Abell 133. {\bf Right}: the inner $4^\circ \times 4^\circ$ around the cluster. The filaments are labeled as described in the text. Inside the X-ray field of view, we show arrows pointing from the cluster center to the end of the filaments overlaid on the X-ray contours. The path from the cluster center to the potential links onto the cosmic web follow along the X-ray filaments.
} \label{fig:filament_idents}
\end{figure*}

We show the combined distribution of galaxies and clusters around Abell 133 at four scales\footnote{We caution that these views are projected rectilinearly from the spherical surface of the sky} in Figure \ref{fig:full_zoomout}. In the broadest view, $20^\circ \times 20^\circ$, we can clearly see that Abell 133 is at a node of the cosmic web. To the west is a highly redshifted filament, while blueshifted structure to the north and east converges on the cluster from the north. Two prominent gaps in the distribution of galaxies exist to the south and east of the cluster; in comparison with Figure \ref{fig:two_xray_images}, these are the directions in which the observed X-ray emission is the weakest and there is no evidence for extended filamentary structure. 

The nearest system to Abell 133 is Abell 114, which is located along the previously mentioned western filament. Abell 114 is approximately 8.2 Mpc from Abell 133, most of which is along the plane of the sky, and should therefore not be contributing to the observed X-ray emission. Abell 114 is, however, only ${\sim}4.5$ Mpc from Abell 2824. These three have been previously associated as being the primary components of a super-cluster, SSCC 41 \citep{2014MNRAS.445.4073C} and are part of the even larger Pisces-Cetus Supercluster \citep{1986ApJ...303...25T,2005MNRAS.364.1387P}. North of Abell 133 is the galaxy group [DZ2015] 351 \citep{2015A&A...578A..61D}, which is ${\sim}9.2$ Mpc away and also lies along a filament of galaxies. 

While the cosmic web can be easily identified at large scales in the 6DF data \citep[as by e.g.,][]{2009MNRAS.399..683J}, the lack of secure redshifts within the inner few degrees makes it difficult to firmly tie the filaments to Abell 133. In the large scale picture presented in Figure \ref{fig:full_zoomout}, several filaments can be identified by eye. In addition to being spatially coherent, they are similarly positioned in redshift (shown in Figure \ref{fig:full_zoomout} by color), which, outside the infall region of Abell 133, means that they are 3-dimensional structures. To identify potential filaments, we used animations of the galaxy distribution, moving through overlapping slices of velocity. In this view, the filaments appear to fall onto the cluster as the center of the velocity slices approaches 0 relative velocity. Our analysis is limited only to those filaments that appear to connect with Abell 133 -- there is significant structure to the west that appears to link other members of the Pisces-Cetus supercluster \citep[e.g.,][]{2005MNRAS.364.1387P}, but the only paths from these structures to Abell 133 require following other filaments already identified.

In total, we highlight 7 filamentary structures that appear to converge on Abell 133, which we show in Figure \ref{fig:filament_idents}. These identifications are not contingent on assigning filament membership probabilities to individual galaxies or on precisely defining the path of the filament in space. Nor do we assert that all seven filaments independently connect to the cluster -- indeed, it seems as they may merge (or bifurcate, depending on one's perspective) as they near the cluster. Rather, these are the only potential visible links between the Cosmic Web and Abell 133 seen in the 6DF survey data, and their apparent connection points to the cluster are in the directions expected from the X-ray observations. 

Working clockwise from north, these filaments are:
\begin{enumerate}
\item a filament positioned almost due North of the cluster
\item a filament to the west / northwest redshifted relative to Abell 133
\item a filament to the southwest
\item a filament to the south / southwest that bends to the east
\item a filament to the southeast that bends to the south
\item a filament to the north that immediately bends east and 
\item a filament to the north / northeast blueshifted relative to Abell 133.
\end{enumerate}

In the right panel of Figure \ref{fig:filament_idents} we show the paths of the filaments as they approach the cluster, but we stop short of connecting them onto Abell 133 due to the lack of comprehensive spectroscopic coverage. Bridging that gap will be the objective of future studies, but for now, we show the path from the cluster center to the ends of the filaments. The alignment -- four potential filamentary links to the north, two to the southwest, and one to the southeast -- is in keeping with the X-ray expectation. 

\section{Discussion}\label{sec:Discussion}

In addition to an investigation into the filaments around a particular cluster, the results presented here should inform forthcoming missions (e.g., {\it AXIS}, {\it Athena}, and {\it Lynx}). We therefore discuss our results in two contexts: the individual filaments that connect to Abell 133 and the lessons we can learn from this analysis for the use in future X-ray observations.

\subsection{The Filaments of Abell 133}

We have not considered the gaseous content of the filaments in this work. This will be covered by our companion paper (Vikhlinin et al, in prep). Nevertheless, from the distribution of galaxies in the large-scale structure, the kinematics of the galaxies near Abell 133, and the positions of the X-ray emission, we can still draw insights into the nature of the cluster filaments.

The nearest cluster, Abell 114, lies along filament 2, which has a relatively high redshift from Abell 133. This filament intersects our {\it Chandra} field of view at the northern border, coming from the west. Filament 7 also intersects the cluster at the northern border and has a similar magnitude of velocity offset from Abell 133, but this offset is of the opposite sign. This pathway extends up to the complex multi-component system of Abell 151 \citep{1996A&A...310...31M}. As the two filaments enter the cluster at the same point and have seemingly mirrored paths in velocity as a function of clustercentric radius, we assume that they are both the same filament. The relatively large velocity offsets (e.g., $2000\ {\rm km}\ {\rm s}^{-1}$ at 20 Mpc) means that, assuming distances from Hubble motions alone, this filament is principally into the plane of the sky. While outside the field of view of our {\it Chandra} data, future X-ray observations may be able to take advantage of the enhanced column density to see this filament and measure its properties.

Presently, we lack any information about the absorption properties of the X-ray filaments, although planned observations with the {\it Hubble Space Telescope} will give us one absorbing sight line \citep{2017hst..prop15198C}. In addition, near filament 3 is TON S180, an AGN at $z = 0.062$ \citep{2009MNRAS.399..683J}. \citet{2004ApJ...614...31B} proposed that observed absorption features in UV observations of TON S180 with HST may be caused by absorption along the supercluster filament we identified. They identified three absorption features at $z = 0.0616$, $z = 0.0620$, and $z= 0.0625$; however, these correspond to the AGN's redshift and absorption redshifted up to ${\sim} 200\ {\rm km }\ {\rm s}^{-1}$ away from the AGN. \citet{2002ApJ...568..120T} also identified these features and included Space Telescope Imaging Spectrograph (STIS) observations in their analysis, but they did not identify any absorption systems beyond that of the AGN. As the nearest filament galaxies are at roughly the redshift of Abell 133, we would expect any filamentary absorption lines to also be at that redshift. Therefore, either Ton S180 is not behind any cosmic filament or the physical conditions of the filament do not cause strong UV absorption features. More comprehensive spectroscopic coverage along the filaments is needed to assess the first option.

One of the few studies providing quantitative predictions for cluster filament connections is that of \citet{2010MNRAS.408.2163A}. Analyzing simulations, they predict that a cluster of mass similar to Abell 133 should have 2-4 filamentary connections to the Cosmic Web. This is in agreement with the results of \citet{2013HEAD...1340101V} and the local distribution of galaxies discussed in Section \ref{sec:optical_ids}. However, we also described up to 7 potential cosmic filaments in Section \ref{sec:CosmicWeb}, which is discrepant both with the X-ray observations, local galaxy distribution, and the theoretical prediction. While this apparent contradiction is simply a result of discussing two different filamentary scales, it is worth considering the observed filamentary network in the context of simulations.

To start, we consider two different filament scales: cosmic and cluster. We define the latter as the bridge between clusters and the cosmic web, while the former is the network of connections between both clusters and filaments. Predictions of 2-4 filaments for Abell 133 and the reported filaments from \citet{2013HEAD...1340101V} are, in this scheme, cluster filaments. \citet{2010MNRAS.408.2163A} required filaments to come within 3 Mpc of the cluster center to be considered a cluster connection, which is slightly larger than the region sampled by the Abell 133 X-ray data.  In simulations of cluster infall regions, \citet{2017MNRAS.464.2027A} show that while the gas distribution out to and slightly beyond $R_{200}$ follows filamentary patterns, {\it these filaments may separate at larger clustercentric radii}.

\citet{2018MNRAS.479..973C} characterized the Cosmic Web in the context of Gaussian random fields. They defined two terms: the connectivity, $\kappa$, and multiplicity, $\mu$. Here $\kappa$ is the number of connections  one peak has to other peaks, while $\mu$ is that value minus the number of bifurcations in linking filaments (such that $\kappa$ traces the number of end points of the filament network around a peak and $\mu$ is the number of starting points at the peak). Observationally, $\kappa$ and $\mu$ correspond to cosmic and cluster filaments, respectively. At redshift $z=0$, \citet{2018MNRAS.479..973C} predict a cluster with mass similar to that of Abell 133 should have connectivity $\kappa \sim 11$, although the probability distribution, $P(\kappa)$, peaks at $\kappa \sim 4$ but has a long tail. While \citet{2018MNRAS.479..973C} do not discuss multiplicity with implicit mass parametrization, probability distributions for multiplicity peak around $\mu \sim 3 - 4$. The increase in the number of cluster (local) to cosmic (large-scale) filaments seen in this work is in keeping with the expectation of \citet{2018MNRAS.479..973C} for bifurcations to increase the number of connections to a node at increasing radii.

While other simulations of large-scale structure lack specific predictions for frequency of filaments, they still provide qualitative expectations for the morphology of filaments and filament bifurcations. \citet{2016MNRAS.457.3024H} identified filaments in the  Illustris simulation. In that work, patterns such as multiple filaments connecting to a halo from one direction (as in the north of Abell 133) are seen. Similar behavior is seen in observations that identify filaments and clusters \citep[e.g.,][]{2014MNRAS.438.3465T,2016MNRAS.461.3896C}.

\subsection{Lessons for Future Observations}
While the X-ray observations discussed here came at considerable expense, future missions should make observations of this caliber much more commonplace. With the planned specifications of {\it eROSITA}, {\it AXIS}, {\it Athena}, and {\it Lynx}, we may, in the next 20 years, have the ability to obtain maps of cosmic filaments for every cluster. Here, we consider how future observing campaigns can be structured to minimize the observational cost of observing filaments, as well as what steps in the analysis can be improved to better identify filaments in X-ray observations.

We first consider the work of \citet{2014MNRAS.437.1909M}. They identified filamentary structure by first subtracting a radially-symmetric surface brightness model and then isolating only high-frequency components in a Radon transformation of the image. By subtracting a symmetric surface brightness profile, they enhanced the excess along the major axis; this would lead to an obvious, linear feature to be isolated by the Radon projection. For any elliptical structure, this technique should have tendency toward finding the major axis of the cluster. The theoretical expectation is that the distribution of matter and galaxies around a cluster can be approximated by an ellipsoid, with two filaments connecting to the clusters along the major axis of this ellipsoid \citep[e.g.,][]{2018MNRAS.479..973C}. These filaments will bifurcate, however, and a bifurcation close enough to the cluster center to be undetectable is typical. Identifying filaments only along the major axis is both minimally helpful (as that is the expectation) and not complete (as it misses bifurcations). Additionally, this method is hampered by the uncertainties in determining the point where the ellipsoid ends and the filament begins. While identifying high-frequency components in Radon space shows the presence of linear structures, this technique requires more development before being applied to cluster observations from future X-ray missions.

In contrast, \citet{2013HEAD...1340101V} removed point sources and identified and subtracted substructure clumps. Due to the faint nature of cosmic filaments, deep observations will be needed to detect them, no matter the telescope. At these depths, the background of X-ray point sources stretches back to at least reionization \citep{2018ApJ...856L..25B}, so point source subtraction -- and high angular resolution -- will be important with future observations. Here, optical and infrared followup can be used to determine if any point sources are associated with the cluster or filaments. \citet{2013HEAD...1340101V} also used wavelet-based detection and subtraction to remove clumps from the X-ray image. While subtracting these structures was needed to identify the filaments, infalling groups have a significant contribution to cluster assembly  \citep{2013ApJ...770...62D}, and so optical observations should be used to identify clumps that should be included in X-ray analyses. Finally, wavelet-based detection and subtraction schemes require accurate background measurement or modeling (see, e.g., \citealt{2017ApJ...848...37C} and \citealt{2017ApJ...835..113L} for applications of this in optical data); over-estimating backgrounds will suppress real signal behind subtracted clumps while under-estimating backgrounds will leave behind false residual structure. As demonstrated in this work, comprehensive spectroscopic followup can be used to verify the presence of detected filaments.

For future X-ray observations of cluster filaments, optical observations of galaxies will be crucial in guiding analysis. In this work, we presented the results of a spectroscopic campaign covering many nights over four years on a 6.5m telescope; this is not scalable with a future where X-ray detections of filaments are routine. Some of this observational cost will be mitigated by current and future redshift surveys \citep[e.g.,][]{2016SPIE.9908E..1OD, 2016arXiv161100036D,2017arXiv170101976E}, which will provide deeper maps of the cosmic web out to larger redshifts, although cluster surveys inside $R_{200}$ will most likely remain a necessity on a case-by-case basis. Deep photometric surveys will also reduce the needed optical observations for an individual cluster, as red sequence analysis can be used to constrain whether clumps are associated with targeted clusters or interlopers \citep[e.g.,][]{2000AJ....120.2148G,2013MsT..........GJ,2014ApJ...785..104R} and photometric information can be used for more efficient spectroscopic targeting. However, for the next generation of X-ray satellites, targeted observations to detect filaments should also include supporting optical observations to maximize the scientific returns.

\section{Summary}
We conducted a spectroscopic campaign to study the distribution of galaxies around Abell 133, where a previous analysis had seen evidence for filaments of the cosmic web seen in X-ray observations, and obtained redshifts for 254 cluster members with IMACS. The distribution of cluster galaxies aligns with what is expected if they are tracer particles of the filaments and the X-ray emission comes from the filaments as well.

\begin{enumerate}
\item In combination with archival redshifts, we compared the positions of cluster members to the projected X-ray emission. We found qualitative agreement between the two samples and a peak in the angular cross-correlation at a significance of ${\sim}5\sigma$. The two filaments with the strongest X-ray emission are matched in the angular distribution of galaxies. The third reported filament, which is weaker in X-rays, shows a loose connection to the third excess in the galaxy distribution, but we cannot directly tie the galaxy distribution to that X-ray filament.
\item From the large-scale distribution of galaxies contained in the 6DF survey, we identified the potential filaments of the cosmic web that can connect to Abell 133. These structures arrive onto the cluster in the directions where X-ray emission is seen.
\item Using the spectroscopic redshifts to determine the kinematic state of the cluster, we found no evidence that there is a significant dynamical disturbance in Abell 133, in contrast to earlier analyses. We found that two previously-identified clumps in the galaxy distribution have no statistically significant kinematic differences from the cluster at large, while the kinematic offset between the cD and the rest of the galaxy population noted by prior works is a consequence of an inaccurate measurement of the cD velocity.
\item Based on the ability of very deep {\it Chandra} observations to see cluster filaments, future X-ray telescopes should be able to see many more filaments. We discussed how future optical observing campaigns should be structured to maximize the scientific returns of these X-ray observations.
\end{enumerate}
\acknowledgments

{\scriptsize We gratefully acknowledge Andrey Kravtsov for thoughtful discussions on the contents of this paper. We thank Mariska Kriek for useful suggestions on supplementary data. The scientific results reported in this article are based on observations made by the {\it Chandra X-ray Observatory}. This paper includes data gathered with the 6.5 meter Magellan Telescopes located at Las Campanas Observatory, Chile. Based on observations obtained with MegaPrime/MegaCam, a joint project of CFHT and CEA/DAPNIA, at the Canada-France-Hawaii Telescope (CFHT) which is operated by the National Research Council (NRC) of Canada, the Institut National des Sciences de l'Univers of the Centre National de la Recherche Scientifique of France, and the University of Hawaii. This research has made use of the NASA/IPAC Extragalactic Database (NED) which is operated by the Jet Propulsion Laboratory, California Institute of Technology, under contract with the National Aeronautics and Space Administration. Based in part on observations at Cerro Tololo Inter-American Observatory, National Optical Astronomy Observatory, which is operated by the Association of Universities for Research in Astronomy (AURA) under a cooperative agreement with the National Science Foundation.}

{\scriptsize This project used public archival data from the Dark Energy Survey (DES). Funding for the DES Projects has been provided by the U.S. Department of Energy, the U.S. National Science Foundation, the Ministry of Science and Education of Spain, the Science and Technology Facilities Council of the United Kingdom, the Higher Education Funding Council for England, the National Center for Supercomputing Applications at the University of Illinois at Urbana-Champaign, the Kavli Institute of Cosmological Physics at the University of Chicago, the Center for Cosmology and Astro-Particle Physics at the Ohio State University, the Mitchell Institute for Fundamental Physics and Astronomy at Texas A\&M University, Financiadora de Estudos e Projetos, Funda{\c c}{\~a}o Carlos Chagas Filho de Amparo {\`a} Pesquisa do Estado do Rio de Janeiro, Conselho Nacional de Desenvolvimento Cient{\'i}fico e Tecnol{\'o}gico and the Minist{\'e}rio da Ci{\^e}ncia, Tecnologia e Inova{\c c}{\~a}o, the Deutsche Forschungsgemeinschaft, and the Collaborating Institutions in the Dark Energy Survey.}

{\scriptsize The Collaborating Institutions are Argonne National Laboratory, the University of California at Santa Cruz, the University of Cambridge, Centro de Investigaciones Energ{\'e}ticas, Medioambientales y Tecnol{\'o}gicas-Madrid, the University of Chicago, University College London, the DES-Brazil Consortium, the University of Edinburgh, the Eidgen{\"o}ssische Technische Hochschule (ETH) Z{\"u}rich,  Fermi National Accelerator Laboratory, the University of Illinois at Urbana-Champaign, the Institut de Ci{\`e}ncies de l'Espai (IEEC/CSIC), the Institut de F{\'i}sica d'Altes Energies, Lawrence Berkeley National Laboratory, the Ludwig-Maximilians Universit{\"a}t M{\"u}nchen and the associated Excellence Cluster Universe, the University of Michigan, the National Optical Astronomy Observatory, the University of Nottingham, The Ohio State University, the OzDES Membership Consortium, the University of Pennsylvania, the University of Portsmouth, SLAC National Accelerator Laboratory, Stanford University, the University of Sussex, and Texas A\&M University.}

\vspace{5mm}
\facilities{Magellan:Baade(IMACS), CXO, Blanco, CFHT}
\software{PyFITS \citep{1999ASPC..172..483B}, 
CarPy \citep{2000ApJ...531..159K, 2003PASP..115..688K}, 
Astropy \citep{2013A&A...558A..33A}}

\appendix

\section{Spectroscopic Completeness} \label{sec:spec_comp}
\begin{figure}
\plotone{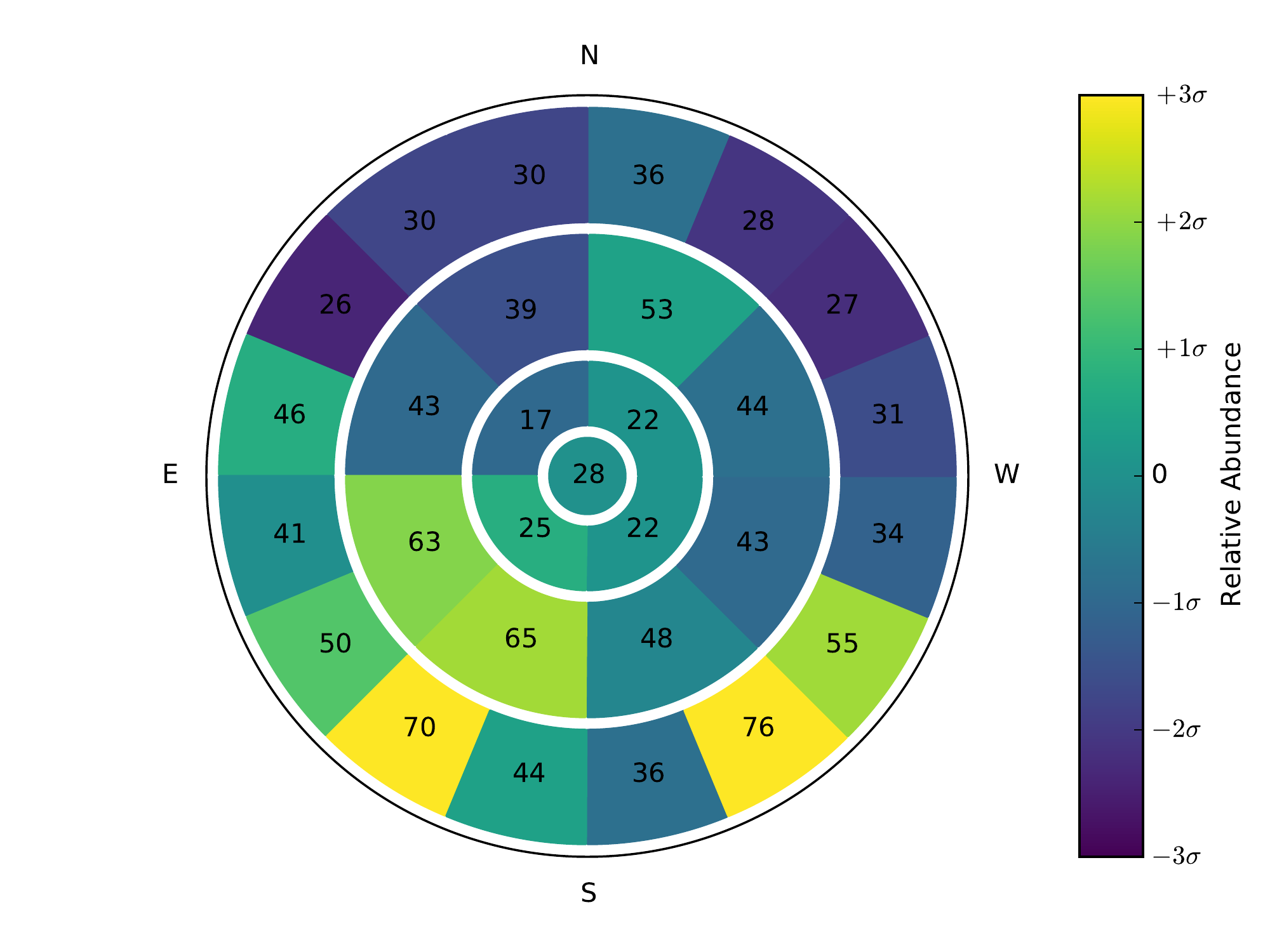}
\caption{Coverage of spectroscopic redshifts around the cluster center among galaxies brighter than $r \leq 22$ mag. Each wedge is labeled according to the number of secure redshifts inside the annular subsection. The coloring is based on the relative abundance of spectroscopic redshifts inside that wedge, where the relative abundance is defined as the deviation from the expected population for a wedge in that annulus, assuming Poissonian statistics. The inner circle has a radius of $200\arcsec$, while the annuli have inner and outer radii of $200\arcsec$ to $500\arcsec$, $500\arcsec$ to $1000\arcsec$, and $1000\arcsec$ to $1500\arcsec$.
 We see a slight bias toward sampling the southern side of the cluster.} \label{fig:radial_spec_coverage}
\end{figure}

Here we discuss the completeness of our spectroscopic observations and any potential biases that may have been introduced.

Our first concern is that one area of the sky having more galaxies around the redshift of the cluster than another area could be due to either filamentary substructure or a relative oversampling of the first region. We therefore consider the distribution of galaxies with good spectroscopic redshifts; in Figure \ref{fig:radial_spec_coverage} we show this distribution in radial wedges. Relative abundances are based on the assumption that each wedge should have $N/W \pm \sqrt{N/W}$ galaxies with redshifts, where N is the total number of known redshifts in the annulus and W is the number of wedges we have divided that annulus into.  While we see a bias toward sampling the south of the cluster, we still detect structure in the N, where we have the worst spatial coverage. Considering Figure \ref{fig:radial_spec_coverage} in the context of Figure \ref{fig:cmd_spec}, we appear to have sampled the galaxy distribution well enough to detect the inherent structure, and the deeper coverage has mostly served to identify interlopers.

\begin{figure}
\plotone{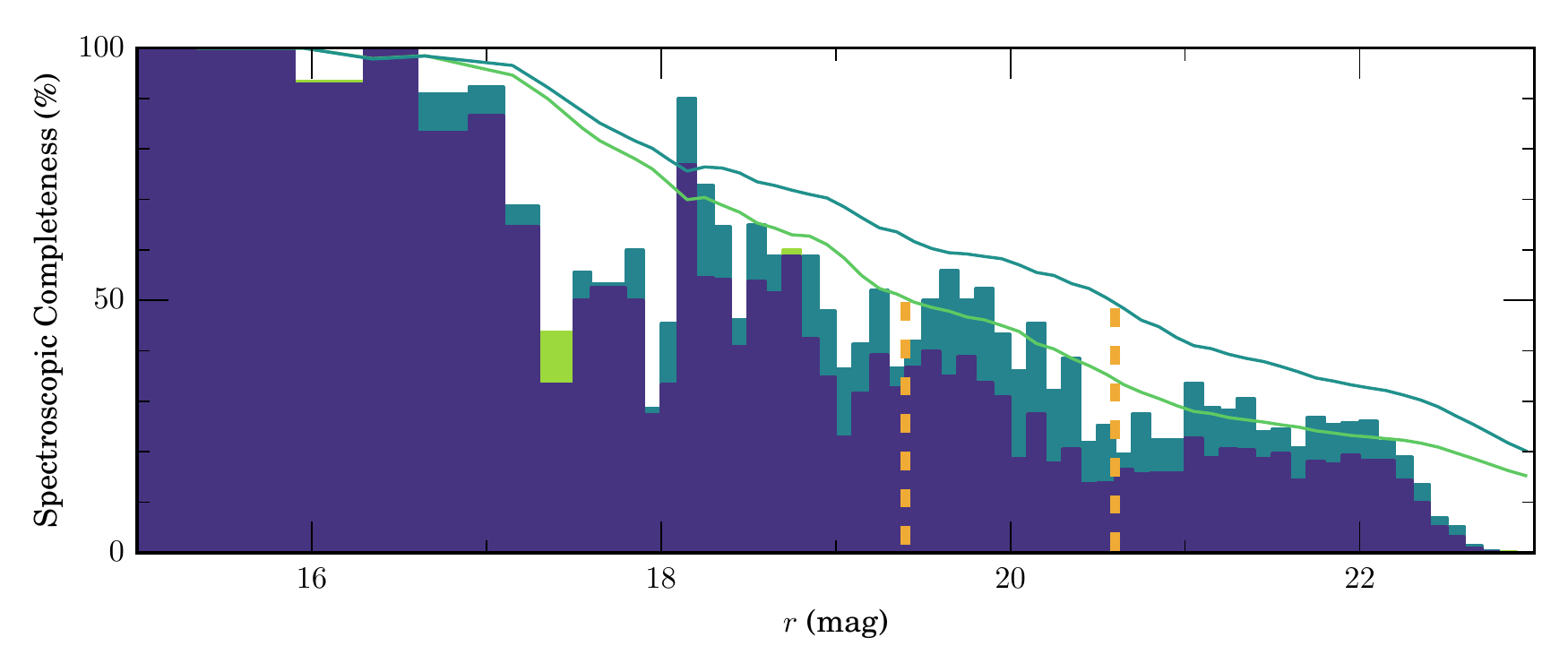}
\caption{Spectroscopic completeness as a function of magnitude for galaxies within $1500\arcsec$ of the cluster center. We consider two populations: all galaxies (green) and blue galaxies (blue, defined as $g - r < 1$). The overlap of the two populations is shown in purple. We show the cumulative fraction for these two populations with the green (all galaxies) and blue (blue galaxies) lines. The magnitudes at which the cumulative fractions reach 50\% are marked by dashed orange lines; these values are $r = 19.4$ and $r = 20.5$ mag for all and blue galaxies, respectively. Bins are sized to contain at least 10 galaxies, and have widths in increments of 0.1 magnitudes.} \label{fig:spectro_completo}
\end{figure}

The second concern is to what depth the samples remain representative of the overall population. As the cluster is expected to be denser and brighter than the filaments, it is important that we were able to sufficiently sample the population of filament galaxies. 
We show in Figure \ref{fig:spectro_completo} the fraction of galaxies within $1500\arcsec$ of the cluster center for which we have obtained redshifts. This is broken into two sets: all galaxies and blue galaxies, where blue galaxies (defined as those with $g - r < 1.0$) include the entire cluster red sequence and everything bluer (see Figure \ref{fig:cmd_spec}). Cumulative 50\% completeness limits are reached at $r = 19.4$ and $r = 20.5$ mag for all and blue galaxies, respectively. We find that we are approximately 100\% complete for galaxies brighter than $r < 17$, ${\sim}50\%$ complete for $17 < r < 20$, and ${\sim} 25\%$ complete from $20 < r < 22$. With a distance modulus of $m - M \sim 37$ and assuming ${\rm M}^* \approx -22.2$ \citep[e.g.,][]{2002PASJ...54..515G}, these bins roughly correspond to 100\% for ${\rm L}/{\rm L}^* \gtrsim 0.1 $, 50\% for $0.1 \gtrsim {\rm L}/{\rm L}^* \gtrsim 0.01 $, and 25\% for $0.01 \gtrsim {\rm L}/{\rm L}^* \gtrsim 0.001 $. While we expect that filaments should have a fainter ${\rm L}^*$ and a lower density of ${\rm L}^*$ galaxies \citep{2005MNRAS.356.1155C}, our survey is deep enough to sample filament galaxies.

Finally, we consider the imprint of our spectroscopic survey on the distribution of galaxies on the sky. Here, we limit ourselves to galaxies brighter than $r < 21$ mag and $r < 19.5$ mag, all with colors bluer than $g-r < 1.0$ (the same color limit used in Figure \ref{fig:spectro_completo}). We use a hexagonal tiling to bin the distribution of galaxies across the X-ray coverage area, and then evaluate the percentage of galaxies with a secure redshift, including those whose redshift comes from the literature. We show our results in Figure \ref{fig:spec_coverage_hexes}. Hexagonal tiles are sized based on the number of galaxies inside the region they cover, such that larger tiles have more galaxies. We also show the contours of the X-ray emission and the boundaries of the X-ray field of view with orange lines.

As can be seen in Figure \ref{fig:spec_coverage_hexes}, our spectroscopic coverage comes to an abrupt end north of the southern border of the X-ray emission. This is a result of our spectroscopic targets being selected from CFHT imaging that did not extend this far to the south. We also have relatively low spectroscopic coverage in the center of the cluster, but this is seen to be due to the large number of potential galaxies in that region. In the southeastern corner of the image, we have relatively low sampling of the potential X-ray filament; in comparison to Figure \ref{fig:specs_on_x}, the part of that filament that does not have any confirmed members also corresponds to the part of the filament with the worst spectroscopic sampling. 
\bibliography{bibliography}

\end{document}